\documentclass[prl, showpacs,twocolumn,nofootinbib,superscriptaddress,floatfix]{revtex4}
\usepackage{amssymb,amsmath}
\usepackage{graphicx}
\usepackage{multirow}
\usepackage{comment}
\usepackage{color}
\usepackage[header,title,page,titletoc]{appendix}
\usepackage{bbm}
\usepackage{amsfonts}%
\usepackage[bookmarks,colorlinks=false]{hyperref}

\begin{document}
\newcommand{\ignore}[1]{}
\def\mc#1{{\mathcal #1}}
\newcommand{\be}{\begin{equation}}
\newcommand{\ee}{\end{equation}}
\newcommand{\ba}{\begin{eqnarray}}
\newcommand{\ea}{\end{eqnarray}}
\newcommand{\order}[1]{\mathcal{O}(#1)}
\newcommand{\cir}[1]{\mathring{#1}}
\graphicspath{{./plots/}}

\title{First observation of the hidden-charm pentaquarks on lattice}

\author{Hanyang Xing}
\affiliation{Institute of Modern Physics, Chinese Academy of Sciences, Lanzhou, 730000, China}
\affiliation{University of Chinese Academy of Sciences, Beijing 100049, China. }

\author{Jian Liang}
\affiliation{Guangdong Provincial Key Laboratory of Nuclear Science, Institute of Quantum Matter, South China Normal University, Guangzhou 510006, China}
\affiliation{Guangdong-Hong Kong Joint Laboratory of Quantum Matter, Southern Nuclear Science Computing Center, South China Normal University, Guangzhou 510006, China}

\author{Liuming Liu}
\email{liuming@impcas.ac.cn}
\affiliation{Institute of Modern Physics, Chinese Academy of Sciences, Lanzhou, 730000, China}
\affiliation{University of Chinese Academy of Sciences, Beijing 100049, China. }

\author{Peng Sun}
\affiliation{Institute of Modern Physics, Chinese Academy of Sciences, Lanzhou, 730000, China}
\affiliation{University of Chinese Academy of Sciences, Beijing 100049, China. }

\author{Yi-Bo Yang}
\affiliation{CAS Key Laboratory of Theoretical Physics, Institute of Theoretical Physics, Chinese Academy of Sciences, Beijing 100190, China}
\affiliation{School of Fundamental Physics and Mathematical Sciences, Hangzhou Institute for Advanced Study, UCAS, Hangzhou 310024, China}
\affiliation{International Centre for Theoretical Physics Asia-Pacific, Beijing/Hangzhou, China}
\affiliation{University of Chinese Academy of Sciences, Beijing 100049, China. }

\begin{abstract}
The s-wave scattering of $\Sigma_c \bar{D}$ and $\Sigma_c \bar{D}^*$ in the $I(J^P) = \frac{1}{2}(\frac{1}{2}^-)$ channel is calculated in lattice QCD using two ensembles with different volumes but the same lattice spacing $a\sim 0.08\mathrm{fm}$ and pion mass $M_\pi \sim 294\mathrm{MeV}$. The scattering amplitudes near threshold are obtained by L\"uscher's finite volume method. We find bound state poles in both $\Sigma_c \bar{D}$ and $\Sigma_c \bar{D}^*$ channels, which are possibly related to the $P_c(4312)$ and $P_c(4440) / P_c(4457)$ pentaquarks observed in experiments. The binding energy is $6(2)(2)$MeV for $\Sigma_c \bar{D}$  and $7(3)(1)$MeV for $\Sigma_c \bar{D}^*$, where the first error is the statistical error and the second is the systematic error due to the lattice artifacts.  

\end{abstract}

\maketitle


\textit{Introduction}: Understanding hadron spectra is a vital part in the understanding of the strong interactions and its underlying theory --- quantum chromodynamics(QCD). Although the QCD theory allows for the existence of any type of colorless hadrons, for a long time the hadrons observed in experiments all fitted into the simple picture of either quark-antiquark pair or three quarks. Only since 2003, new types of hadrons were started to be observed in experiments, triggered tremendous interest and effort in the study of hadron spectra. One of the most important discoveries was the hidden-charm pentaquarks $P_c(4450)$ and $P_c(4380)$ reported by the LHCb collaboration in 2015~\cite{LHCb:2015yax}. A later analysis based on a data sample that is an order of magnitude larger shows that  the $P_c(4450)$ splits into two structures $P_c(4440)$ and $P_c(4457)$, and a third narrow peak $P_c(4312)$ emerges~\cite{LHCb:2019kea}. Numerous theoretical investigations on the nature of the $P_c$ states followed these discoveries. Several theoretical interpretations have been proposed, including hadronic molecules~\cite{Du:2021fmf, Chen:2019bip,Chen:2019asm,Guo:2019fdo,Liu:2019tjn,Guo:2019kdc, Xiao:2019mst,Xiao:2019aya,Meng:2019ilv,Xiao:2019gjd,Yamaguchi:2019seo,Liu:2019zvb,Lin:2019qiv,Wang:2019ato,Burns:2019iih,Du:2019pij,Wang:2019spc}, compact pentaquark states~\cite{Ali:2019npk,Wang:2019got,Cheng:2019obk} and  hadrocharmonia~\cite{Eides:2019tgv,Ferretti:2018ojb}. It is readily observed that the mass of $P_c(4312)$ is close to the $\Sigma_c\bar{D}$ thresholds while the $P_c(4440)$ and $P_c(4457)$ are close to the $\Sigma_c\bar{D}^*$ threshold. The molecule interpretation naturally explains all of the three narrow $P_c$ states as spin multiplets of the $\Sigma_c \bar{D}^{(*)}$ bound states.
 In most of the literatures, the $P_c(4312)$ is explained as a $J^P = \frac{1}{2}^-$ $\Sigma_c\bar{D}$ bound state, while the $P_c(4440)$ and $P_c(4457)$ are the $\Sigma_c\bar{D}^*$ bound states with quantum numbers $\frac{1}{2}^-$ and $\frac{3}{2}^-$(or $\frac{3}{2}^-$ and $\frac{1}{2}^-$) respectively~\cite{Du:2021fmf,Chen:2019bip,Chen:2019asm,Liu:2019tjn,Xiao:2019mst,Xiao:2019aya,Meng:2019ilv,Yamaguchi:2019seo,Liu:2019zvb,Lin:2019qiv,Wang:2019ato,Du:2019pij,Wang:2019spc}. 

The study of the $P_c$ pentaquarks from first-principle lattice QCD calculation is still lacking. There is a pioneering lattice calculation of the nucleon-$J/\psi$ and nucleon-$\eta_c$ scattering aiming at searching for the $P_c$ pentaquarks, but found no strong indication for a resonance or bound state in these scattering channels~\cite{Skerbis:2018lew}. In light of the phenomenological studies, we study the $\Sigma_c\bar{D}^{(*)}$ interactions and  investigate whether they can form bound states in this work. We focus on the $J^P = \frac{1}{2}^-$ channels and ignore the effects of other coupled channels such as $J/\psi p$, $\eta_c p$ and $\Lambda_c\bar{D}^{(*)}$. In Ref.~\cite{Guo:2019kdc}, the authors calculated the probabilities of finding the $J/\psi p$, $\Lambda_c \bar{D}^{(*)}$ and $\Sigma_c\bar{D}^{(*)}$ components inside the $P_c$ states in the framework of the effective-range expansion and resonance compositeness relations, and found that the weight of $\Sigma_c\bar{D}^{(*)}$ is much larger than $J/\psi p$ and $\Lambda_c\bar{D}^{(*)}$. Other coupled-channel studies also indicate that $\Sigma_c\bar{D}^{(*)}$ is the dominant channel~\cite{Xiao:2019aya,Du:2021fmf,Yalikun:2021bfm}. 
As the first attempt to calculate the $\Sigma_c\bar{D}^{(*)}$ interactions in lattice QCD, we do not consider the couplings of other channels and find out whether any structure will emerge in the single channel scattering.

A well established method to study scattering processes in lattice QCD is the L\"uscher's finite volume method~\cite{Luscher:1990ux}, which relates the finite-volume spectrum of a two-particle system to the scattering parameters of the two particles in the infinite volume. The finite-volume spectrum can be calculated rather straightforwardly in lattice QCD, then the scattering amplitudes can be obtained through L\"uscher's method. Resonances and bound states appear as poles in the scattering amplitudes. In this letter we calculate the $J^P = \frac{1}{2}^-$ $\Sigma_c\bar{D}$ and $\Sigma_c\bar{D}^*$ scattering via the L\"uscher's method, and find bound state poles in both channels. This is the first time the signals of the hidden-charm pentaquarks are observed in first-principle lattice calculations. 

In the following, we first present the lattice setup and the finite-volume spectrum, and then give the results of the scattering analysis.

\textit{Lattice setup}:
The results presented in this letter are based on two ensembles of gauge configurations with 2+1 dynamical quark flavors. The gauge action is tree-level Symanzik-improved while in the fermion section we use the Shekholeslami-Wohlert action~\cite{Sheikholeslami:1985ij} with tree-level tadpole improvement. The two ensembles have the same parameters except for the volume. The lattice spacing is $~\sim$ 0.080~fm which is determined by gradient flow~\cite{Borsanyi:2012zs}. The light quark mass is heavier than the physical value, corresponding to a pion mass $M_\pi \sim 294 \mathrm{MeV}$. The charm quark mass is tuned to produce the physical spin-averaged mass of $\eta_c$ and $J/\Psi$, i.e. $\frac{1}{4}M_{\eta_c} + \frac{3}{4}M_{J/\Psi}$. The spacial volume is $L=32$ and $48$ for the two ensembles respectively. The ensemble with $L=32$ has been used to study the semileptonic decays of charmed-strange baryons~\cite{Zhang:2021oja}.
The parameters of the ensembles are listed in TABLE~\ref{Table:configs}. 

\begin{table}[tb]
\begin{tabular}{cccccccc}
\hline
  ID &$a(\mathrm{fm})$  &$am_l$    &$am_s$   &$M_{\pi}$(MeV)          &$L^3 \times T$  & $N_{\mathrm{conf.}}$\\
\hline
 L32        & 0.0805(14) &-0.2295  &-0.2050  &294.5(0.9) &$32^3 \times 96$ &    371       \\
 L48        & 0.0803(05)  &-0.2295  &-0.2050   &294.2(0.5) &$48^3 \times 96$ &   201        \\
\hline
\end{tabular}
\caption{Parameters of the ensembles. The listed parameters are the lattice spacing $a$, bare quark masses for the light($m_l$) and strange($m_s$) quark, the pion mass $M_\pi$,  the lattice volume $L^3\times T$ and the number of configurations $N_{\mathrm{conf.}}$.}
\label{Table:configs}
\end{table}

The masses of the relevant single particles---$D$, $D^*$ and $\Sigma_c$---are computed from the correlation functions of the corresponding single particle interpolating operators. The dispersion relation $E^2 = m_0^2 + c^2p^2$  is investigated by calculating the single-particle energy of $D$, $D^*$ and $\Sigma_c$ at the five lowest momenta on lattice(in units of $2\pi/L$): $(0,0,0)$, $(0,0,1)$, $(0,1,1)$, $(1,1,1)$, $(0,0,2)$. For each particle, we fit the five energies at the five momenta to the dispersion relation and get the parameters $m_0$ and $c$. The results are collected in TABLE~\ref{Table:SingleParticleSpectra}. The values of $c^2$ generally deviate from the physical value 1 within statistical uncertainties due to the lattice artifacts mainly from the charm quark. We will take care of this issue in the scattering analysis later. 

The distillation quark smearing method ~\cite{HadronSpectrum:2009krc} is used to compute the quark propagators. This method enables us to greatly improve the precision with affordable cost and conveniently compute the correlation functions of many interpolating operators. More details of the distillation method, interpolators, single particle energies and fits to the dispersion relation can be found in the supplemental materials~\cite{supplemental}.

\begin{table*}[tb]
\begin{tabular}{|c|c|c|c|c|c|c|}
\hline
  \multirow{2}{*}{} & \multicolumn{2}{c|}{$D$} & \multicolumn{2}{c|}{$D^* $} &\multicolumn{2}{c|}{$ \Sigma_c$}  \\
  \cline{2-7} 
   &$m_0$(GeV) &$c^2$  &$m_0$(GeV)  &$c^2$  &$m_0$(GeV)  & $c^2$ \\
 \hline
 L32 &1.8920(7) &0.936(5) &1.9984(14) &0.920(10) &2.4755(24) &0.999(13)  \\
 \hline
 L48 &1.8971(4) &0.933(4) &2.0017(8) &0.920(8) &2.4723(16) &0.938(22) \\
 \hline
\end{tabular}
\caption{Fit results of the dispersion relation for $D$, $D^*$ and $\Sigma_c$. }
\label{Table:SingleParticleSpectra}
\end{table*}

\textit{Finite-volume spectrum}:
In order to get the finite-volume energies, one first needs to construct a set of interpolating operators. We are interested in the $\Sigma_c\bar{D}^{(*)}$ interactions in the $I(J^P) = \frac{1}{2}({\frac{1}{2}}^-)$ channel. Three $\Sigma_c\bar{D}$ operators and two $\Sigma_c\bar{D}^*$ operators with different momentum combinations are used to cover the energy range near the thresholds that we are interested. These operators can be written as  
\ba
\mathcal{O}^{\Sigma_c\bar{D}}_{\mathbf{p_1}, \mathbf{p_2}}  &=&\sum_{\alpha, \mathbf{p_1}, \mathbf{p_2}} C_{\alpha, \mathbf{p_1}, \mathbf{p_2}} \big( \sqrt{\frac{2}{3}} \Sigma_{c, \alpha}^{++} (\mathbf{p_{1}})D^{-} (\mathbf{p_{2}}) \nonumber \\ 
&& - \sqrt{\frac{1}{3}}\Sigma_{c, \alpha}^{+}(\mathbf{p_{1}}) \bar{D}^0(\mathbf{p_{2}}) \big),\\
\mathcal{O}^{\Sigma_c\bar{D}^*}_{\mathbf{p_1}, \mathbf{p_2}}&=& \sum_{\alpha, k, \mathbf{p_1}, \mathbf{p_2}} C_{\alpha, k, \mathbf{p_1}, \mathbf{p_2}}  \big( \sqrt{\frac{2}{3}} \Sigma_{c, \alpha}^{++}(\mathbf{p_1}) D_k^{*-}(\mathbf{p_2}) \nonumber \\ 
&& - \sqrt{\frac{1}{3}}\Sigma_{c, \alpha}^{+}(\mathbf{p_1}) \bar{D}_k^{*0}(\mathbf{p_2})\big),
\ea

where $\alpha$ represents the Dirac index of the $\Sigma_c$ baryon operators, $k$ is the vector index of the vector meson $\bar{D}^*$. In this work we use only the operators with zero total momentum $ \mathbf{p_1} + \mathbf{p_2} = \mathbf{0}$. In a periodic cubic box of size $L$, the momentum is quantized as $(2\pi/L)\cdot \mathbf{n}$, with $\mathbf{n} \in Z^3$. The three $\Sigma_c\bar{D}$ operators have $|\mathbf{p_{1,2}}| = 0$,  $2\pi/L$ and $(2\pi/L)\cdot \sqrt{2}$ respectively, while the two $\Sigma_c\bar{D}^*$ operators have $|\mathbf{p_{1,2}}|=0$ and  $2\pi/L$ respectively. The coefficients $C_{\alpha, \mathbf{p_1}, \mathbf{p_2}}$ and  $C_{\alpha, k, \mathbf{p_1}, \mathbf{p_2}}$ are chosen so that the the operators transform in the $G_1^-$ irreducible representation of the cubic group which corresponds to $J^P = \frac{1}{2}^-$ in the continuum, and their values can be found in the supplementary materials~\cite{supplemental}. We follow the operator construction method described in Ref.~\cite{Prelovsek:2016iyo}. 

We compute the matrix of the correlation functions of the five operators described above
 \be C_{ij}(t) = \sum_{t_{src}} \langle\mathcal{O}_i(t+t_{src}) \mathcal{O}_j^\dagger(t_{src})\rangle. \ee 
 Solving the generalized eigenvalue problem(GEVP)
\be
C(t) v^n(t) = \lambda^n(t) C(t_0) v^n(t),
\ee
the energies can be extracted from the time dependence of the eigenvalues $\lambda^n(t)$ ~\cite{Luscher:1990ck}. We choose $t_0=4$  and fit the eigenvalues to a two-exponential form $\lambda^n(t) = (1-A_n) e^{-E_n(t-t_0)} + A_n e^{-E_n^\prime(t-t_0)}$, where the fit parameters are $A_n$, $E_n$ and $E_n^\prime$. The energies obtained from the fits are plotted in FIG.~\ref{Figure:FV_energy} for the two ensembles, together with the non-interacting two-particle energies
\be
E^{\mathrm{free}} = \sqrt{m_1^2 +  \mathbf{p_1}^2} +  \sqrt{m_2^2 + \mathbf{p_2}^2},
\label{Eq:freeE}
\ee where $m_1$ and $m_2$ are the masses of the two particles. There are tiny differences between the single particle masses of the two ensembles. The non-interacting energies in FIG.~\ref{Figure:FV_energy} are calculated with $m_1$ and $m_2$ taking the values of the L48 ensemble. The data points of the L32 ensemble are shifted accordingly to show the correct gaps between the finite-volume energies and the non-interacting energies for this ensemble. 
It is observed that the finite-volume energies are generally below the non-interacting energies, indicating rather strong attractive interactions. We also observe that the mixing between the $\Sigma_c \bar{D}$ and $\Sigma_c \bar{D}^*$ operators is negligible. 
If we perform the GEVP analysis with the three $\Sigma_c\bar{D}$ operators (or the two $\Sigma_c \bar{D}^*$ operators),  the energies we get agree perfectly with the three red points (or the two blue points) shown in FIG.~\ref{Figure:FV_energy} for each volume, and the values are given in the supplementary materials~\cite{supplemental}. Therefore, we associate the red (blue) data points to the  $\Sigma_c\bar{D}$ ($\Sigma_c\bar{D}^*$) channel and perform single channel scattering analysis for $\Sigma_c\bar{D}$ and $\Sigma_c\bar{D}^*$ separately. 

\begin{figure}
\includegraphics[width =0.5 \textwidth]{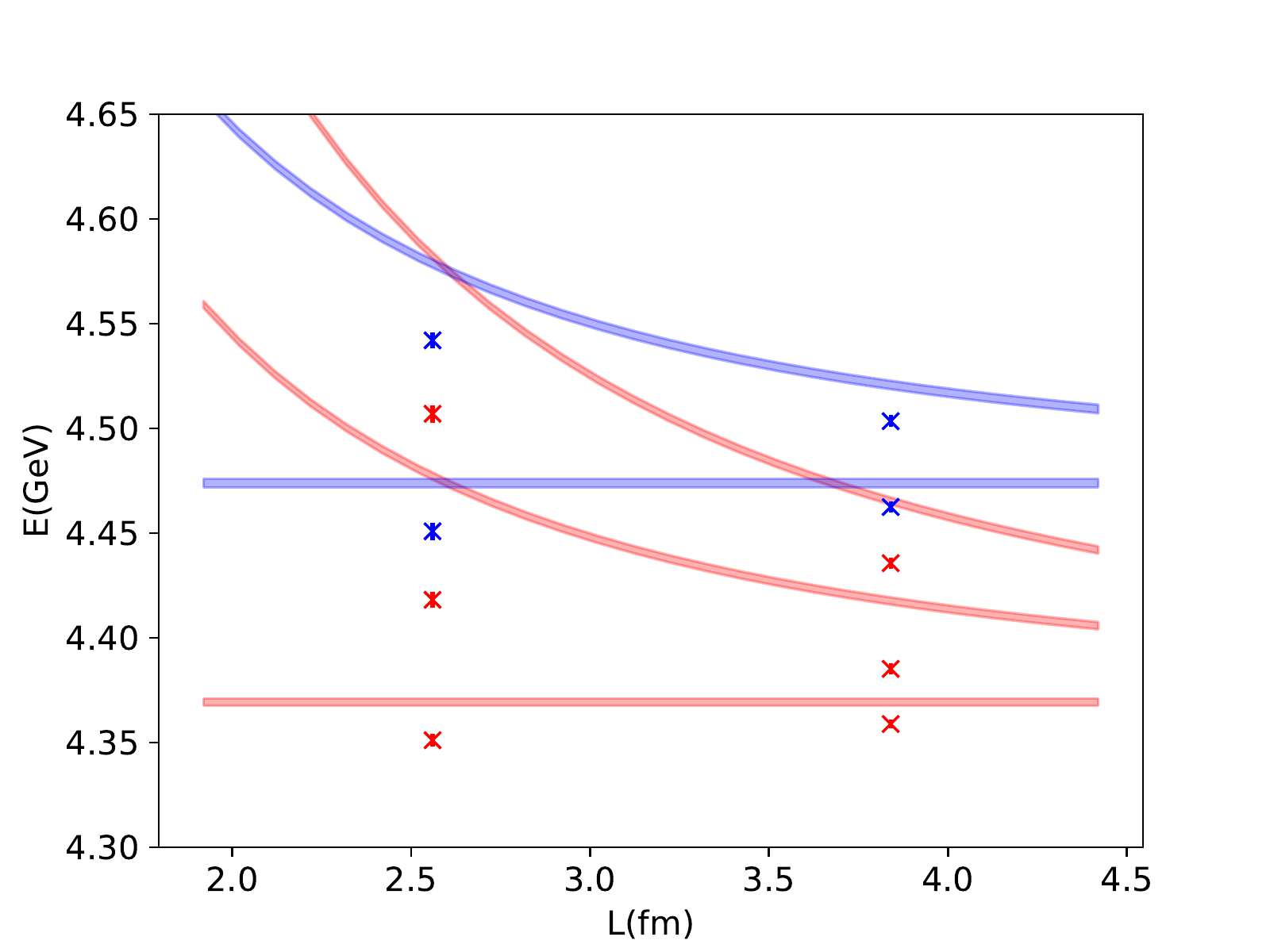}
\caption{The red (blue) data points are the finite-volume energies associated with the $\Sigma_c \bar{D}$ ($\Sigma_c \bar{D}^*$) operators as described in the text. The red (blue) bands indicate the free energies of the non-interacting $\Sigma_c \bar{D}$ ($\Sigma_c \bar{D}^*$) threshold with different momenta. The width of the bands represent the statistical uncertainties.}
\label{Figure:FV_energy}
\end{figure}

\textit{Scattering analysis}:
L\"uscher's finite volume method provides a direct relation between the energy eigenvalues of a two-particle system in a finite box and the scattering phase shift of the two particles in the infinite volume. In the non-interacting case, the energy of the two particles takes the form of Eq.~\ref{Eq:freeE}. In the presence of interactions, the finite-volume energies of the two-particle system will shift from the free energies. For the single channel s-wave scattering, L\"uscher's formula that relates the finite-volume energy and the infinite volume scattering phase shift reads
\be
p\cot \delta_0(p) = \frac{2}{L\sqrt{\pi}} \mathcal{Z}_{00}(1; q^2),
\label{Eq:Luscher}
\ee
where $\delta_0$ is the s-wave scattering phase shift, $p$ is the scattering momentum obtained from finite-volume energy $E = \sqrt{m_1^2 + p^2} +  \sqrt{m_2^2 + p^2}$, and $q = \frac{pL}{2\pi}$. The zeta-function $\mathcal{Z}_{00}(1; q^2)$ can be evaluated numerically once $q^2$ is given.

The scattering amplitude with partial wave $l$ and total angular momentum $J$ can be written as
\be
t_l^{(J)} \sim \frac{1}{p\cot\delta_l^{(J)}-ip}.
\ee
The hadrons appear as poles in the scattering amplitude. A bound state corresponds to a pole on the real axis below the threshold (and therefore $p^2 < 0$) in the first Riemann sheet, which is $p=i|p_B|$. $p_B$ denotes the value of $p$ where the bound state pole occurs. 

For the $J^{P} = \frac{1}{2}^-$ $\Sigma_c \bar{D}$ scattering, the partial wave $l=0$.  We parametrize the scattering amplitude near threshold with the effective range expansion up to $\mathcal{O}(p^2)$
\be
p \cot \delta_0 = \frac{1}{a_0} + \frac{1}{2} r_0 p^2. 
\label{Eq:ERE}
\ee
The values of $p \cot \delta_0$ is evaluated from Eq.~\ref{Eq:Luscher} and plotted in FIG.~\ref{Figure:pcot_SigmacD} as a function of $p^2$. We use the five energies associate with the $\Sigma_c\bar{D}$ operators (the red points  in FIG.~\ref{Figure:FV_energy})  below the $\Sigma_c \bar{D}^*$ threshold for the analysis. 

As we mentioned before, the continuum dispersion relation for the single particles $D$, $D^*$ and $\Sigma_c$ is not perfectly preserved due to the lattice artifacts. The lattice energies can deviate from the continuum values. Considering that the L\"uscher's formula is based on the continuum dispersion relation, we  shift the two-particle finite volume energies with respect to the differences of the continuum and lattice single particle energies: 
\be
E_{p}^{\prime} = E_{p} + (M_{D(p)}^{\mathrm{cont.}} + M_{\Sigma_c(p)}^{\mathrm{cont.}}) - (M_{D(p)}^{\mathrm{lat.}} + M_{\Sigma_c(p)}^{\mathrm{lat.}}),
\label{Eq:shiftE}
\ee
where $E_{p}$ is the two-particle energy that has dominant contribution from the operator $\mathcal{O}^{\Sigma_c \bar{D}}_{\mathbf{p_1}, \mathbf{p_2}}$ with $|\mathbf{p_{1,2}}| = p$. $M_{D(p)}^{\mathrm{cont.}}$ and $M_{\Sigma_c(p)}^{\mathrm{cont.}}$ are the energies of $D$ and $\Sigma_c$ at momentum $p$  calculated from the continuum dispersion relation, $M_{D(p)}^{\mathrm{lat.}}$ and $M_{\Sigma_c(p)}^{\mathrm{lat.}}$ are the corresponding energies computed on lattice. This strategy is argued and applied in the studies of charmed meson scattering with similar lattice spacing as we use in this work~\cite{Padmanath:2022cvl, Prelovsek:2020eiw, Piemonte:2019cbi}. As a comparison, we also did the scattering analysis with the original two-particle energies obtained from GEVP. The differences will be taken as the systematic uncertainty caused by lattice artifacts. In FIG.~\ref{Figure:pcot_SigmacD}, the upper plot shows the results using the shifted energies while the lower plot shows the results using the original energies. The grey band in each plot represents the fit of the data points to the effective range expansion Eq.~\ref{Eq:ERE}, and the red curve is $ip=-|p|$ versus $p^2$. The $p^2$ value corresponds to the intersection of the grey band and the red curve is where the bound state pole occurs, which will be denoted as $p_B^2$. Note that there are two poles appearing in our results, one is close to the threshold and the other one is far below the threshold (not shown in FIG.~\ref{Figure:pcot_SigmacD}). We take the one that is close to the threshold as the physical one. The value of $p_B^2$ and the fitted parameters $a_0$ and $r_0$ are listed in TABLE~\ref{tab:EREfit_SigmacD}. The binding energy $E_B = M_D + M_{\Sigma_c} - (\sqrt{M_D^2 + p_B^2} + \sqrt{M_{\Sigma_c}^2 + p_B^2})$ is also presented in the table. 

\begin{figure}
\includegraphics[width =0.5 \textwidth]{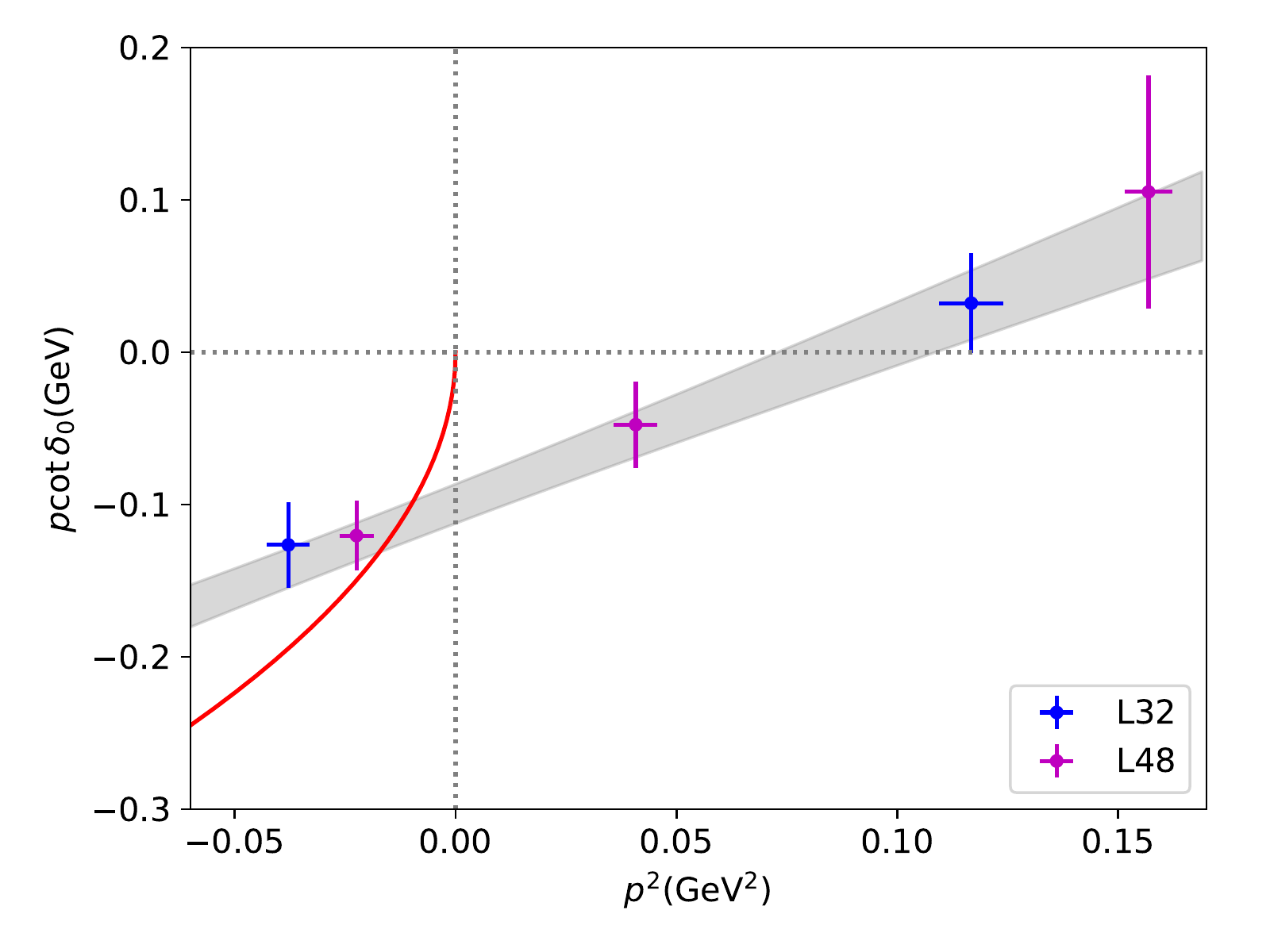}
\includegraphics[width =0.5 \textwidth]{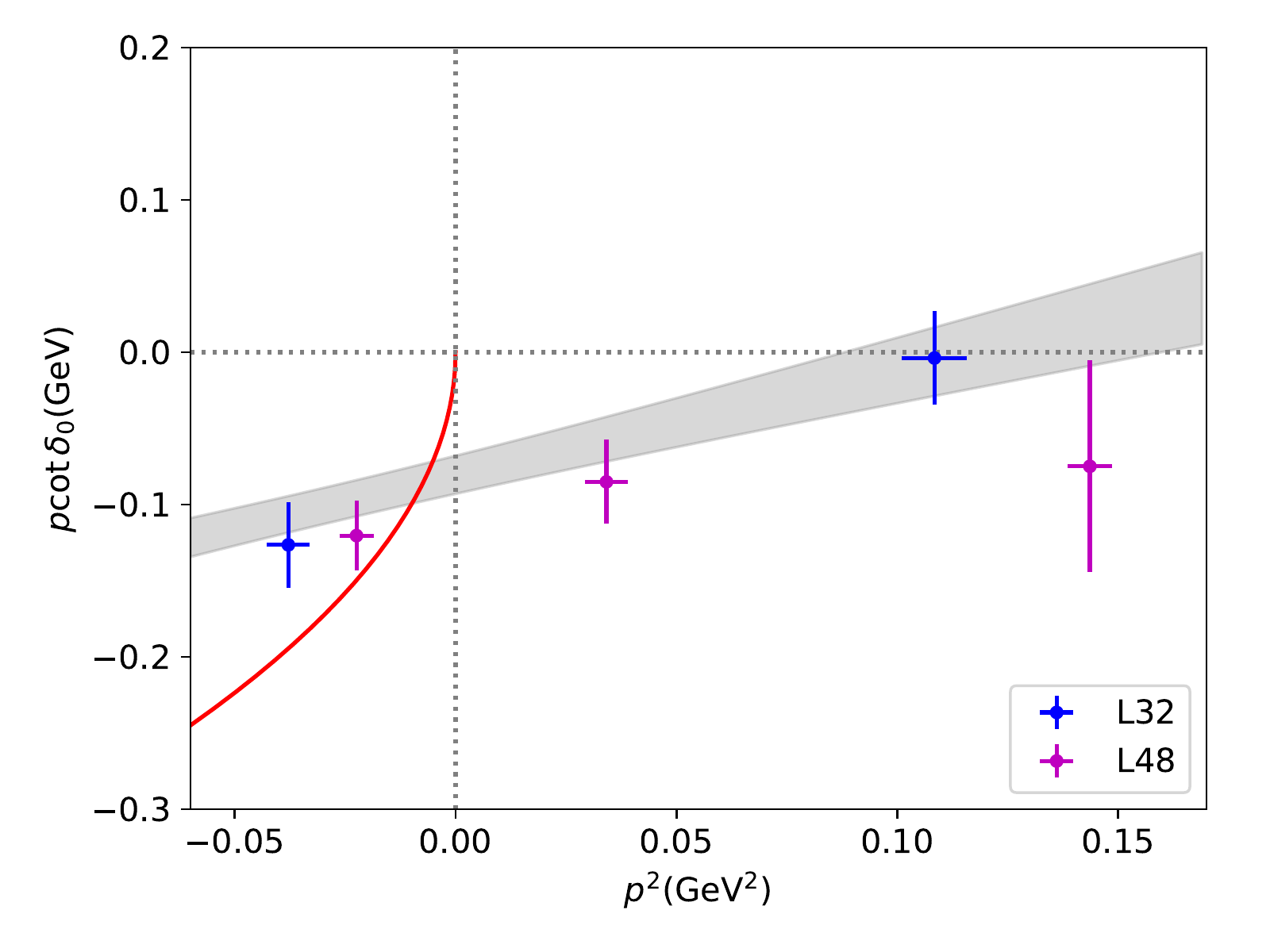}
\caption{Fit of the $p\cot \delta_0$ to the effective range expansion for the $\Sigma_c \bar{D}$ scattering. The red curve is $ip=-|p|$ versus $p^2$. The bound state pole occurs at the intersection of the fitted band (the grey band) and the red curve.}
\label{Figure:pcot_SigmacD}
\end{figure}

\begin{table}[t!]
\begin{tabular}{|c|c|c|c|c|c|}
\hline
 & $a_0$ (fm) & $r_0$ (fm)  &$p_B^2$ (GeV$^2$) &$E_B$ (MeV) &$\chi^2/$d.o.f. \\
 \hline
 fit1 &-2.0(3) &0.46(6) &-0.013(4) &6(2) &0.28 \\
 \hline
 fit2 &-2.5(4) &0.29(6) &-0.007(2) &4(1) &1.00 \\
 \hline
\end{tabular}
\caption{Fit results for the $\Sigma_c\bar{D}$ scattering. The scattering length $a_0$ and effective range $r_0$ are obtained by fitting the $p \cot\delta_0$ to the effective range expansion form. $p_B^2$ is the $p^2$ value that the bound state pole occurs. $E_B$ is the binding energy. ``fit1" and ``fit2" are the results using the shifted energies and the original energies respectively as explained in the text.}
\label{tab:EREfit_SigmacD}
\end{table}

For the $\Sigma_c \bar{D}^*$ scattering, we did the scattering analysis similarly using the four energies associated with the $\Sigma_c \bar{D}^*$ operators (the blue points in FIG.~\ref{Figure:FV_energy}), and also found bound state pole in the scattering amplitude. The results are presented in FIG.~\ref{Figure:pcot_SigmacDstar} and TABLE~\ref{tab:EREfit_SigmacDstar}. 
\begin{figure}
\includegraphics[width =0.5 \textwidth]{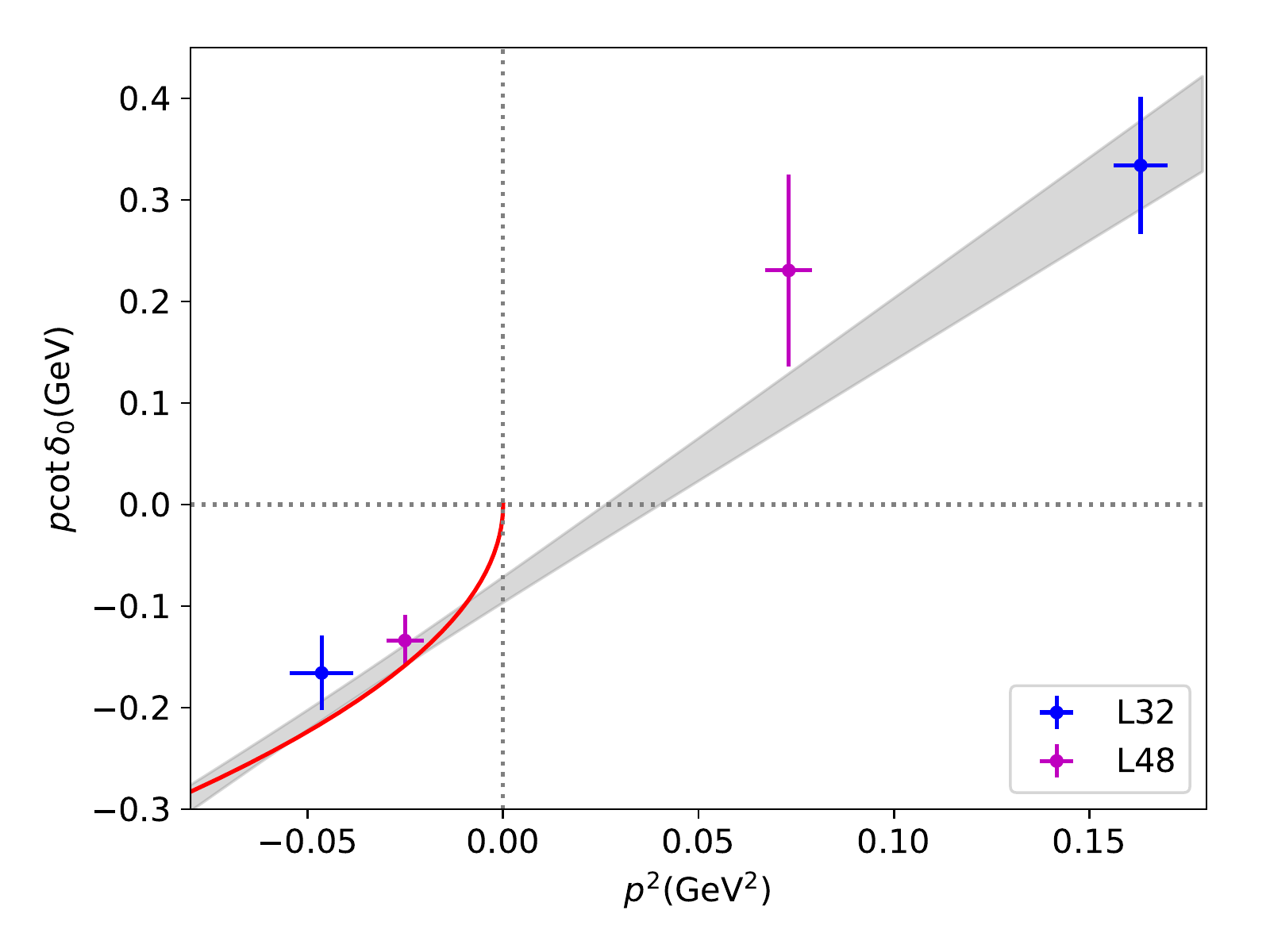}
\includegraphics[width =0.5 \textwidth]{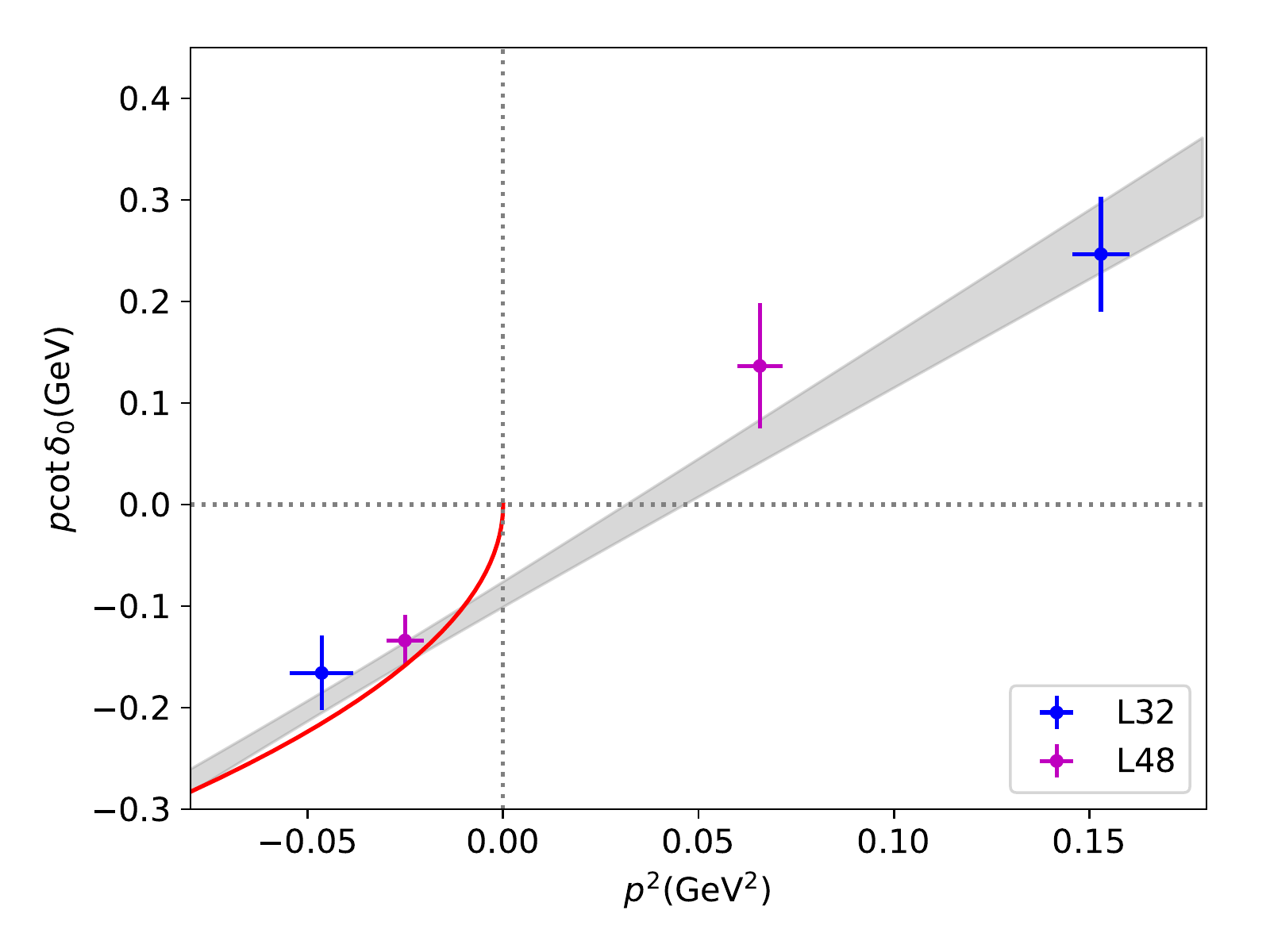}
\caption{Same as FIG.~\ref{Figure:pcot_SigmacD}, but for $\Sigma_c \bar{D}^*$ scattering.}
\label{Figure:pcot_SigmacDstar}
\end{figure}

\begin{table}[t!]
\begin{tabular}{|c|c|c|c|c|c|}
\hline
 & $a_0$ (fm) & $r_0$ (fm)  &$p_B^2$ (GeV$^2$) &$E_B$ (MeV) &$\chi^2/$d.o.f. \\
 \hline
fit1 &-2.3(5) &1.01(8) &-0.016(7) &7(3) &1.6 \\
 \hline
fit2 &-2.2(4) &0.91(7) &-0.017(6) &8(3) &1.5 \\
 \hline
\end{tabular}
\caption{Same as TABLE~\ref{tab:EREfit_SigmacD}, but for $\Sigma_c\bar{D}^*$ scattering.}
\label{tab:EREfit_SigmacDstar}
\end{table}

Our final results of the scattering length, effective range and binding energy for the s-wave $\Sigma_c\bar{D}$ and $\Sigma_c\bar{D}^*$ scattering are:
\ba
a_0(\Sigma_c\bar{D}) &=& -2.0(3)(5) \mathrm{fm}, \nonumber \\
r_0(\Sigma_c\bar{D}) &=& 0.46(6)(17) \mathrm{fm}, \\
E_B(\Sigma_c\bar{D}) &=& 6(2)(2) \mathrm{MeV}, \nonumber
\ea

\ba
a_0(\Sigma_c\bar{D}^*) &=& -2.3(5)(1) \mathrm{fm}, \nonumber \\
r_0(\Sigma_c\bar{D}^*) &=& 1.01(8)(10) \mathrm{fm}, \\
E_B(\Sigma_c\bar{D}^*) &=& 7(3)(1) \mathrm{MeV}, \nonumber
\ea
where the first error is the statistical error and the second error is the systematic error due to the lattice artifacts.

\textit{Summary and discussions}: We study the s-wave scattering of $\Sigma_c\bar{D}$ and $\Sigma_c\bar{D}^*$ with $I(J^P) = \frac{1}{2}(\frac{1}{2}^-$) in lattice QCD at the pion mass $M_\pi = 294$MeV. Bound state poles are found in both channels. The $\Sigma_c\bar{D}$ bound state is possibly related to the $P_c(4312)$ and the binding energy we obtain is 6(2)(2)MeV, which is close to the physical value $\sim 9$MeV. The $\Sigma_c\bar{D}^*$ bound state could be $P_c(4440)$ or $P_c(4457)$. The binding energy from our calculation 7(3)(1)MeV is close to the physical value for $P_c(4457)$. But with the current precision and without the information of $\frac{3}{2}^-$ $\Sigma_c\bar{D}^*$ scattering, we are not able to tell the $\frac{1}{2}^-$ $\Sigma_c\bar{D}^*$ bound state we observe is $P_c(4440)$ or $P_c(4457)$. Future studies with the following improvements are desired to further clarify the properties of the $P_c$ states: including coupled channels, performing the calculation at a smaller lattice spacing and at the physical pion mass to really connect the lattice results to the experimental values. 

\textit{Acknowledgements}: The gauge configurations used in this study are generated as a joint effort of the CLQCD collaboration. This work used the computation resources of the Southern Nuclear Science Computing Center(SNSC) and the Gansu Advanced Computing Center. HX and LL thanks the support from  the CAS Interdisciplinary Innovation Team program,  the Strategic Priority Research Program of Chinese Academy of Sciences with Grant No.\ XDB34030301. YY is supported by Strategic Priority Research Program of Chinese Academy of Sciences, Grant No. XDC01040100, XDB34030303, XDPB1, and also a NSFC-DFG joint grant under grant No. 12061131006 and SCHA 458/22. This work is supported in part by the National Science Foundation of China (NSFC) under Projects No.12175279 and No.12175073. 

\bibliographystyle{apsrev}
\bibliography{reference}

\begin{thebibliography}{37}
\expandafter\ifx\csname natexlab\endcsname\relax\def\natexlab#1{#1}\fi
\expandafter\ifx\csname bibnamefont\endcsname\relax
  \def\bibnamefont#1{#1}\fi
\expandafter\ifx\csname bibfnamefont\endcsname\relax
  \def\bibfnamefont#1{#1}\fi
\expandafter\ifx\csname citenamefont\endcsname\relax
  \def\citenamefont#1{#1}\fi
\expandafter\ifx\csname url\endcsname\relax
  \def\url#1{\texttt{#1}}\fi
\expandafter\ifx\csname urlprefix\endcsname\relax\def\urlprefix{URL }\fi
\providecommand{\bibinfo}[2]{#2}
\providecommand{\eprint}[2][]{\url{#2}}

\bibitem[{\citenamefont{Aaij et~al.}(2015)}]{LHCb:2015yax}
\bibinfo{author}{\bibfnamefont{R.}~\bibnamefont{Aaij}} \bibnamefont{et~al.}
  (\bibinfo{collaboration}{LHCb}), \bibinfo{journal}{Phys. Rev. Lett.}
  \textbf{\bibinfo{volume}{115}}, \bibinfo{pages}{072001}
  (\bibinfo{year}{2015}), \eprint{1507.03414}.

\bibitem[{\citenamefont{Aaij et~al.}(2019)}]{LHCb:2019kea}
\bibinfo{author}{\bibfnamefont{R.}~\bibnamefont{Aaij}} \bibnamefont{et~al.}
  (\bibinfo{collaboration}{LHCb}), \bibinfo{journal}{Phys. Rev. Lett.}
  \textbf{\bibinfo{volume}{122}}, \bibinfo{pages}{222001}
  (\bibinfo{year}{2019}), \eprint{1904.03947}.

\bibitem[{\citenamefont{Du et~al.}(2021)\citenamefont{Du, Baru, Guo, Hanhart,
  Mei\ss{}ner, Oller, and Wang}}]{Du:2021fmf}
\bibinfo{author}{\bibfnamefont{M.-L.} \bibnamefont{Du}},
  \bibinfo{author}{\bibfnamefont{V.}~\bibnamefont{Baru}},
  \bibinfo{author}{\bibfnamefont{F.-K.} \bibnamefont{Guo}},
  \bibinfo{author}{\bibfnamefont{C.}~\bibnamefont{Hanhart}},
  \bibinfo{author}{\bibfnamefont{U.-G.} \bibnamefont{Mei\ss{}ner}},
  \bibinfo{author}{\bibfnamefont{J.~A.} \bibnamefont{Oller}}, \bibnamefont{and}
  \bibinfo{author}{\bibfnamefont{Q.}~\bibnamefont{Wang}},
  \bibinfo{journal}{JHEP} \textbf{\bibinfo{volume}{08}}, \bibinfo{pages}{157}
  (\bibinfo{year}{2021}), \eprint{2102.07159}.

\bibitem[{\citenamefont{Chen et~al.}(2019{\natexlab{a}})\citenamefont{Chen,
  Chen, and Zhu}}]{Chen:2019bip}
\bibinfo{author}{\bibfnamefont{H.-X.} \bibnamefont{Chen}},
  \bibinfo{author}{\bibfnamefont{W.}~\bibnamefont{Chen}}, \bibnamefont{and}
  \bibinfo{author}{\bibfnamefont{S.-L.} \bibnamefont{Zhu}},
  \bibinfo{journal}{Phys. Rev. D} \textbf{\bibinfo{volume}{100}},
  \bibinfo{pages}{051501} (\bibinfo{year}{2019}{\natexlab{a}}),
  \eprint{1903.11001}.

\bibitem[{\citenamefont{Chen et~al.}(2019{\natexlab{b}})\citenamefont{Chen,
  Sun, Liu, and Zhu}}]{Chen:2019asm}
\bibinfo{author}{\bibfnamefont{R.}~\bibnamefont{Chen}},
  \bibinfo{author}{\bibfnamefont{Z.-F.} \bibnamefont{Sun}},
  \bibinfo{author}{\bibfnamefont{X.}~\bibnamefont{Liu}}, \bibnamefont{and}
  \bibinfo{author}{\bibfnamefont{S.-L.} \bibnamefont{Zhu}},
  \bibinfo{journal}{Phys. Rev. D} \textbf{\bibinfo{volume}{100}},
  \bibinfo{pages}{011502} (\bibinfo{year}{2019}{\natexlab{b}}),
  \eprint{1903.11013}.

\bibitem[{\citenamefont{Guo et~al.}(2019)\citenamefont{Guo, Jing, Mei\ss{}ner,
  and Sakai}}]{Guo:2019fdo}
\bibinfo{author}{\bibfnamefont{F.-K.} \bibnamefont{Guo}},
  \bibinfo{author}{\bibfnamefont{H.-J.} \bibnamefont{Jing}},
  \bibinfo{author}{\bibfnamefont{U.-G.} \bibnamefont{Mei\ss{}ner}},
  \bibnamefont{and} \bibinfo{author}{\bibfnamefont{S.}~\bibnamefont{Sakai}},
  \bibinfo{journal}{Phys. Rev. D} \textbf{\bibinfo{volume}{99}},
  \bibinfo{pages}{091501} (\bibinfo{year}{2019}), \eprint{1903.11503}.

\bibitem[{\citenamefont{Liu et~al.}(2019)\citenamefont{Liu, Pan, Peng,
  S\'anchez~S\'anchez, Geng, Hosaka, and Pavon~Valderrama}}]{Liu:2019tjn}
\bibinfo{author}{\bibfnamefont{M.-Z.} \bibnamefont{Liu}},
  \bibinfo{author}{\bibfnamefont{Y.-W.} \bibnamefont{Pan}},
  \bibinfo{author}{\bibfnamefont{F.-Z.} \bibnamefont{Peng}},
  \bibinfo{author}{\bibfnamefont{M.}~\bibnamefont{S\'anchez~S\'anchez}},
  \bibinfo{author}{\bibfnamefont{L.-S.} \bibnamefont{Geng}},
  \bibinfo{author}{\bibfnamefont{A.}~\bibnamefont{Hosaka}}, \bibnamefont{and}
  \bibinfo{author}{\bibfnamefont{M.}~\bibnamefont{Pavon~Valderrama}},
  \bibinfo{journal}{Phys. Rev. Lett.} \textbf{\bibinfo{volume}{122}},
  \bibinfo{pages}{242001} (\bibinfo{year}{2019}), \eprint{1903.11560}.

\bibitem[{\citenamefont{Guo and Oller}(2019)}]{Guo:2019kdc}
\bibinfo{author}{\bibfnamefont{Z.-H.} \bibnamefont{Guo}} \bibnamefont{and}
  \bibinfo{author}{\bibfnamefont{J.~A.} \bibnamefont{Oller}},
  \bibinfo{journal}{Phys. Lett. B} \textbf{\bibinfo{volume}{793}},
  \bibinfo{pages}{144} (\bibinfo{year}{2019}), \eprint{1904.00851}.

\bibitem[{\citenamefont{Xiao et~al.}(2019{\natexlab{a}})\citenamefont{Xiao,
  Huang, Dong, Geng, and Chen}}]{Xiao:2019mst}
\bibinfo{author}{\bibfnamefont{C.-J.} \bibnamefont{Xiao}},
  \bibinfo{author}{\bibfnamefont{Y.}~\bibnamefont{Huang}},
  \bibinfo{author}{\bibfnamefont{Y.-B.} \bibnamefont{Dong}},
  \bibinfo{author}{\bibfnamefont{L.-S.} \bibnamefont{Geng}}, \bibnamefont{and}
  \bibinfo{author}{\bibfnamefont{D.-Y.} \bibnamefont{Chen}},
  \bibinfo{journal}{Phys. Rev. D} \textbf{\bibinfo{volume}{100}},
  \bibinfo{pages}{014022} (\bibinfo{year}{2019}{\natexlab{a}}),
  \eprint{1904.00872}.

\bibitem[{\citenamefont{Xiao et~al.}(2019{\natexlab{b}})\citenamefont{Xiao,
  Nieves, and Oset}}]{Xiao:2019aya}
\bibinfo{author}{\bibfnamefont{C.~W.} \bibnamefont{Xiao}},
  \bibinfo{author}{\bibfnamefont{J.}~\bibnamefont{Nieves}}, \bibnamefont{and}
  \bibinfo{author}{\bibfnamefont{E.}~\bibnamefont{Oset}},
  \bibinfo{journal}{Phys. Rev. D} \textbf{\bibinfo{volume}{100}},
  \bibinfo{pages}{014021} (\bibinfo{year}{2019}{\natexlab{b}}),
  \eprint{1904.01296}.

\bibitem[{\citenamefont{Meng et~al.}(2019)\citenamefont{Meng, Wang, Wang, and
  Zhu}}]{Meng:2019ilv}
\bibinfo{author}{\bibfnamefont{L.}~\bibnamefont{Meng}},
  \bibinfo{author}{\bibfnamefont{B.}~\bibnamefont{Wang}},
  \bibinfo{author}{\bibfnamefont{G.-J.} \bibnamefont{Wang}}, \bibnamefont{and}
  \bibinfo{author}{\bibfnamefont{S.-L.} \bibnamefont{Zhu}},
  \bibinfo{journal}{Phys. Rev. D} \textbf{\bibinfo{volume}{100}},
  \bibinfo{pages}{014031} (\bibinfo{year}{2019}), \eprint{1905.04113}.

\bibitem[{\citenamefont{Xiao et~al.}(2019{\natexlab{c}})\citenamefont{Xiao,
  Nieves, and Oset}}]{Xiao:2019gjd}
\bibinfo{author}{\bibfnamefont{C.~W.} \bibnamefont{Xiao}},
  \bibinfo{author}{\bibfnamefont{J.}~\bibnamefont{Nieves}}, \bibnamefont{and}
  \bibinfo{author}{\bibfnamefont{E.}~\bibnamefont{Oset}},
  \bibinfo{journal}{Phys. Lett. B} \textbf{\bibinfo{volume}{799}},
  \bibinfo{pages}{135051} (\bibinfo{year}{2019}{\natexlab{c}}),
  \eprint{1906.09010}.

\bibitem[{\citenamefont{Yamaguchi et~al.}(2020)\citenamefont{Yamaguchi,
  Garc\'\i{}a-Tecocoatzi, Giachino, Hosaka, Santopinto, Takeuchi, and
  Takizawa}}]{Yamaguchi:2019seo}
\bibinfo{author}{\bibfnamefont{Y.}~\bibnamefont{Yamaguchi}},
  \bibinfo{author}{\bibfnamefont{H.}~\bibnamefont{Garc\'\i{}a-Tecocoatzi}},
  \bibinfo{author}{\bibfnamefont{A.}~\bibnamefont{Giachino}},
  \bibinfo{author}{\bibfnamefont{A.}~\bibnamefont{Hosaka}},
  \bibinfo{author}{\bibfnamefont{E.}~\bibnamefont{Santopinto}},
  \bibinfo{author}{\bibfnamefont{S.}~\bibnamefont{Takeuchi}}, \bibnamefont{and}
  \bibinfo{author}{\bibfnamefont{M.}~\bibnamefont{Takizawa}},
  \bibinfo{journal}{Phys. Rev. D} \textbf{\bibinfo{volume}{101}},
  \bibinfo{pages}{091502} (\bibinfo{year}{2020}), \eprint{1907.04684}.

\bibitem[{\citenamefont{Liu et~al.}(2021)\citenamefont{Liu, Wu,
  S\'anchez~S\'anchez, Valderrama, Geng, and Xie}}]{Liu:2019zvb}
\bibinfo{author}{\bibfnamefont{M.-Z.} \bibnamefont{Liu}},
  \bibinfo{author}{\bibfnamefont{T.-W.} \bibnamefont{Wu}},
  \bibinfo{author}{\bibfnamefont{M.}~\bibnamefont{S\'anchez~S\'anchez}},
  \bibinfo{author}{\bibfnamefont{M.~P.} \bibnamefont{Valderrama}},
  \bibinfo{author}{\bibfnamefont{L.-S.} \bibnamefont{Geng}}, \bibnamefont{and}
  \bibinfo{author}{\bibfnamefont{J.-J.} \bibnamefont{Xie}},
  \bibinfo{journal}{Phys. Rev. D} \textbf{\bibinfo{volume}{103}},
  \bibinfo{pages}{054004} (\bibinfo{year}{2021}), \eprint{1907.06093}.

\bibitem[{\citenamefont{Lin and Zou}(2019)}]{Lin:2019qiv}
\bibinfo{author}{\bibfnamefont{Y.-H.} \bibnamefont{Lin}} \bibnamefont{and}
  \bibinfo{author}{\bibfnamefont{B.-S.} \bibnamefont{Zou}},
  \bibinfo{journal}{Phys. Rev. D} \textbf{\bibinfo{volume}{100}},
  \bibinfo{pages}{056005} (\bibinfo{year}{2019}), \eprint{1908.05309}.

\bibitem[{\citenamefont{Wang et~al.}(2019)\citenamefont{Wang, Meng, and
  Zhu}}]{Wang:2019ato}
\bibinfo{author}{\bibfnamefont{B.}~\bibnamefont{Wang}},
  \bibinfo{author}{\bibfnamefont{L.}~\bibnamefont{Meng}}, \bibnamefont{and}
  \bibinfo{author}{\bibfnamefont{S.-L.} \bibnamefont{Zhu}},
  \bibinfo{journal}{JHEP} \textbf{\bibinfo{volume}{11}}, \bibinfo{pages}{108}
  (\bibinfo{year}{2019}), \eprint{1909.13054}.

\bibitem[{\citenamefont{Burns and Swanson}(2019)}]{Burns:2019iih}
\bibinfo{author}{\bibfnamefont{T.~J.} \bibnamefont{Burns}} \bibnamefont{and}
  \bibinfo{author}{\bibfnamefont{E.~S.} \bibnamefont{Swanson}},
  \bibinfo{journal}{Phys. Rev. D} \textbf{\bibinfo{volume}{100}},
  \bibinfo{pages}{114033} (\bibinfo{year}{2019}), \eprint{1908.03528}.

\bibitem[{\citenamefont{Du et~al.}(2020)\citenamefont{Du, Baru, Guo, Hanhart,
  Mei\ss{}ner, Oller, and Wang}}]{Du:2019pij}
\bibinfo{author}{\bibfnamefont{M.-L.} \bibnamefont{Du}},
  \bibinfo{author}{\bibfnamefont{V.}~\bibnamefont{Baru}},
  \bibinfo{author}{\bibfnamefont{F.-K.} \bibnamefont{Guo}},
  \bibinfo{author}{\bibfnamefont{C.}~\bibnamefont{Hanhart}},
  \bibinfo{author}{\bibfnamefont{U.-G.} \bibnamefont{Mei\ss{}ner}},
  \bibinfo{author}{\bibfnamefont{J.~A.} \bibnamefont{Oller}}, \bibnamefont{and}
  \bibinfo{author}{\bibfnamefont{Q.}~\bibnamefont{Wang}},
  \bibinfo{journal}{Phys. Rev. Lett.} \textbf{\bibinfo{volume}{124}},
  \bibinfo{pages}{072001} (\bibinfo{year}{2020}), \eprint{1910.11846}.

\bibitem[{\citenamefont{Wang et~al.}(2020)\citenamefont{Wang, Xiao, Chen, Liu,
  Liu, and Zhu}}]{Wang:2019spc}
\bibinfo{author}{\bibfnamefont{G.-J.} \bibnamefont{Wang}},
  \bibinfo{author}{\bibfnamefont{L.-Y.} \bibnamefont{Xiao}},
  \bibinfo{author}{\bibfnamefont{R.}~\bibnamefont{Chen}},
  \bibinfo{author}{\bibfnamefont{X.-H.} \bibnamefont{Liu}},
  \bibinfo{author}{\bibfnamefont{X.}~\bibnamefont{Liu}}, \bibnamefont{and}
  \bibinfo{author}{\bibfnamefont{S.-L.} \bibnamefont{Zhu}},
  \bibinfo{journal}{Phys. Rev. D} \textbf{\bibinfo{volume}{102}},
  \bibinfo{pages}{036012} (\bibinfo{year}{2020}), \eprint{1911.09613}.

\bibitem[{\citenamefont{Ali and Parkhomenko}(2019)}]{Ali:2019npk}
\bibinfo{author}{\bibfnamefont{A.}~\bibnamefont{Ali}} \bibnamefont{and}
  \bibinfo{author}{\bibfnamefont{A.~Y.} \bibnamefont{Parkhomenko}},
  \bibinfo{journal}{Phys. Lett. B} \textbf{\bibinfo{volume}{793}},
  \bibinfo{pages}{365} (\bibinfo{year}{2019}), \eprint{1904.00446}.

\bibitem[{\citenamefont{Wang}(2020)}]{Wang:2019got}
\bibinfo{author}{\bibfnamefont{Z.-G.} \bibnamefont{Wang}},
  \bibinfo{journal}{Int. J. Mod. Phys. A} \textbf{\bibinfo{volume}{35}},
  \bibinfo{pages}{2050003} (\bibinfo{year}{2020}), \eprint{1905.02892}.

\bibitem[{\citenamefont{Cheng and Liu}(2019)}]{Cheng:2019obk}
\bibinfo{author}{\bibfnamefont{J.-B.} \bibnamefont{Cheng}} \bibnamefont{and}
  \bibinfo{author}{\bibfnamefont{Y.-R.} \bibnamefont{Liu}},
  \bibinfo{journal}{Phys. Rev. D} \textbf{\bibinfo{volume}{100}},
  \bibinfo{pages}{054002} (\bibinfo{year}{2019}), \eprint{1905.08605}.

\bibitem[{\citenamefont{Eides et~al.}(2020)\citenamefont{Eides, Petrov, and
  Polyakov}}]{Eides:2019tgv}
\bibinfo{author}{\bibfnamefont{M.~I.} \bibnamefont{Eides}},
  \bibinfo{author}{\bibfnamefont{V.~Y.} \bibnamefont{Petrov}},
  \bibnamefont{and} \bibinfo{author}{\bibfnamefont{M.~V.}
  \bibnamefont{Polyakov}}, \bibinfo{journal}{Mod. Phys. Lett. A}
  \textbf{\bibinfo{volume}{35}}, \bibinfo{pages}{2050151}
  (\bibinfo{year}{2020}), \eprint{1904.11616}.

\bibitem[{\citenamefont{Ferretti et~al.}(2019)\citenamefont{Ferretti,
  Santopinto, Naeem~Anwar, and Bedolla}}]{Ferretti:2018ojb}
\bibinfo{author}{\bibfnamefont{J.}~\bibnamefont{Ferretti}},
  \bibinfo{author}{\bibfnamefont{E.}~\bibnamefont{Santopinto}},
  \bibinfo{author}{\bibfnamefont{M.}~\bibnamefont{Naeem~Anwar}},
  \bibnamefont{and} \bibinfo{author}{\bibfnamefont{M.~A.}
  \bibnamefont{Bedolla}}, \bibinfo{journal}{Phys. Lett. B}
  \textbf{\bibinfo{volume}{789}}, \bibinfo{pages}{562} (\bibinfo{year}{2019}),
  \eprint{1807.01207}.

\bibitem[{\citenamefont{Skerbis and Prelovsek}(2019)}]{Skerbis:2018lew}
\bibinfo{author}{\bibfnamefont{U.}~\bibnamefont{Skerbis}} \bibnamefont{and}
  \bibinfo{author}{\bibfnamefont{S.}~\bibnamefont{Prelovsek}},
  \bibinfo{journal}{Phys. Rev. D} \textbf{\bibinfo{volume}{99}},
  \bibinfo{pages}{094505} (\bibinfo{year}{2019}), \eprint{1811.02285}.

\bibitem[{\citenamefont{Yalikun et~al.}(2021)\citenamefont{Yalikun, Lin, Guo,
  Kamiya, and Zou}}]{Yalikun:2021bfm}
\bibinfo{author}{\bibfnamefont{N.}~\bibnamefont{Yalikun}},
  \bibinfo{author}{\bibfnamefont{Y.-H.} \bibnamefont{Lin}},
  \bibinfo{author}{\bibfnamefont{F.-K.} \bibnamefont{Guo}},
  \bibinfo{author}{\bibfnamefont{Y.}~\bibnamefont{Kamiya}}, \bibnamefont{and}
  \bibinfo{author}{\bibfnamefont{B.-S.} \bibnamefont{Zou}},
  \bibinfo{journal}{Phys. Rev. D} \textbf{\bibinfo{volume}{104}},
  \bibinfo{pages}{094039} (\bibinfo{year}{2021}), \eprint{2109.03504}.

\bibitem[{\citenamefont{Luscher}(1991)}]{Luscher:1990ux}
\bibinfo{author}{\bibfnamefont{M.}~\bibnamefont{Luscher}},
  \bibinfo{journal}{Nucl. Phys. B} \textbf{\bibinfo{volume}{354}},
  \bibinfo{pages}{531} (\bibinfo{year}{1991}).

\bibitem[{\citenamefont{Sheikholeslami and
  Wohlert}(1985)}]{Sheikholeslami:1985ij}
\bibinfo{author}{\bibfnamefont{B.}~\bibnamefont{Sheikholeslami}}
  \bibnamefont{and} \bibinfo{author}{\bibfnamefont{R.}~\bibnamefont{Wohlert}},
  \bibinfo{journal}{Nucl. Phys. B} \textbf{\bibinfo{volume}{259}},
  \bibinfo{pages}{572} (\bibinfo{year}{1985}).

\bibitem[{\citenamefont{Borsanyi et~al.}(2012)}]{Borsanyi:2012zs}
\bibinfo{author}{\bibfnamefont{S.}~\bibnamefont{Borsanyi}}
  \bibnamefont{et~al.}, \bibinfo{journal}{JHEP} \textbf{\bibinfo{volume}{09}},
  \bibinfo{pages}{010} (\bibinfo{year}{2012}), \eprint{1203.4469}.

\bibitem[{\citenamefont{Zhang et~al.}(2022)}]{Zhang:2021oja}
\bibinfo{author}{\bibfnamefont{Q.-A.} \bibnamefont{Zhang}}
  \bibnamefont{et~al.}, \bibinfo{journal}{Chin. Phys. C}
  \textbf{\bibinfo{volume}{46}}, \bibinfo{pages}{011002}
  (\bibinfo{year}{2022}), \eprint{2103.07064}.

\bibitem[{\citenamefont{Peardon et~al.}(2009)\citenamefont{Peardon, Bulava,
  Foley, Morningstar, Dudek, Edwards, Joo, Lin, Richards, and
  Juge}}]{HadronSpectrum:2009krc}
\bibinfo{author}{\bibfnamefont{M.}~\bibnamefont{Peardon}},
  \bibinfo{author}{\bibfnamefont{J.}~\bibnamefont{Bulava}},
  \bibinfo{author}{\bibfnamefont{J.}~\bibnamefont{Foley}},
  \bibinfo{author}{\bibfnamefont{C.}~\bibnamefont{Morningstar}},
  \bibinfo{author}{\bibfnamefont{J.}~\bibnamefont{Dudek}},
  \bibinfo{author}{\bibfnamefont{R.~G.} \bibnamefont{Edwards}},
  \bibinfo{author}{\bibfnamefont{B.}~\bibnamefont{Joo}},
  \bibinfo{author}{\bibfnamefont{H.-W.} \bibnamefont{Lin}},
  \bibinfo{author}{\bibfnamefont{D.~G.} \bibnamefont{Richards}},
  \bibnamefont{and} \bibinfo{author}{\bibfnamefont{K.~J.} \bibnamefont{Juge}}
  (\bibinfo{collaboration}{Hadron Spectrum}), \bibinfo{journal}{Phys. Rev. D}
  \textbf{\bibinfo{volume}{80}}, \bibinfo{pages}{054506}
  (\bibinfo{year}{2009}), \eprint{0905.2160}.

\bibitem[{sup(2022)}]{supplemental}
\bibinfo{journal}{See the supplemental materials for further details on the
  interpolating operators, the computation and analysis of the single-particle
  and two-particle correlation functions.}  (\bibinfo{year}{2022}).

\bibitem[{\citenamefont{Prelovsek et~al.}(2017)\citenamefont{Prelovsek,
  Skerbis, and Lang}}]{Prelovsek:2016iyo}
\bibinfo{author}{\bibfnamefont{S.}~\bibnamefont{Prelovsek}},
  \bibinfo{author}{\bibfnamefont{U.}~\bibnamefont{Skerbis}}, \bibnamefont{and}
  \bibinfo{author}{\bibfnamefont{C.~B.} \bibnamefont{Lang}},
  \bibinfo{journal}{JHEP} \textbf{\bibinfo{volume}{01}}, \bibinfo{pages}{129}
  (\bibinfo{year}{2017}), \eprint{1607.06738}.

\bibitem[{\citenamefont{Luscher and Wolff}(1990)}]{Luscher:1990ck}
\bibinfo{author}{\bibfnamefont{M.}~\bibnamefont{Luscher}} \bibnamefont{and}
  \bibinfo{author}{\bibfnamefont{U.}~\bibnamefont{Wolff}},
  \bibinfo{journal}{Nucl. Phys. B} \textbf{\bibinfo{volume}{339}},
  \bibinfo{pages}{222} (\bibinfo{year}{1990}).

\bibitem[{\citenamefont{Padmanath and Prelovsek}(2022)}]{Padmanath:2022cvl}
\bibinfo{author}{\bibfnamefont{M.}~\bibnamefont{Padmanath}} \bibnamefont{and}
  \bibinfo{author}{\bibfnamefont{S.}~\bibnamefont{Prelovsek}},
  \bibinfo{journal}{Phys. Rev. Lett.} \textbf{\bibinfo{volume}{129}},
  \bibinfo{pages}{032002} (\bibinfo{year}{2022}), \eprint{2202.10110}.

\bibitem[{\citenamefont{Prelovsek et~al.}(2021)\citenamefont{Prelovsek,
  Collins, Mohler, Padmanath, and Piemonte}}]{Prelovsek:2020eiw}
\bibinfo{author}{\bibfnamefont{S.}~\bibnamefont{Prelovsek}},
  \bibinfo{author}{\bibfnamefont{S.}~\bibnamefont{Collins}},
  \bibinfo{author}{\bibfnamefont{D.}~\bibnamefont{Mohler}},
  \bibinfo{author}{\bibfnamefont{M.}~\bibnamefont{Padmanath}},
  \bibnamefont{and} \bibinfo{author}{\bibfnamefont{S.}~\bibnamefont{Piemonte}},
  \bibinfo{journal}{JHEP} \textbf{\bibinfo{volume}{06}}, \bibinfo{pages}{035}
  (\bibinfo{year}{2021}), \eprint{2011.02542}.

\bibitem[{\citenamefont{Piemonte et~al.}(2019)\citenamefont{Piemonte, Collins,
  Mohler, Padmanath, and Prelovsek}}]{Piemonte:2019cbi}
\bibinfo{author}{\bibfnamefont{S.}~\bibnamefont{Piemonte}},
  \bibinfo{author}{\bibfnamefont{S.}~\bibnamefont{Collins}},
  \bibinfo{author}{\bibfnamefont{D.}~\bibnamefont{Mohler}},
  \bibinfo{author}{\bibfnamefont{M.}~\bibnamefont{Padmanath}},
  \bibnamefont{and}
  \bibinfo{author}{\bibfnamefont{S.}~\bibnamefont{Prelovsek}},
  \bibinfo{journal}{Phys. Rev. D} \textbf{\bibinfo{volume}{100}},
  \bibinfo{pages}{074505} (\bibinfo{year}{2019}), \eprint{1905.03506}.

\end{thebibliography}

\clearpage
\setcounter{equation}{0}
\setcounter{figure}{0}
\setcounter{table}{0}

\begin{center}
  {\large \bf Supplemental material}
\end{center}

In this supplemental material we provide additional information about our study of the hidden-charm pentaquarks. 
\maketitle

\section{Interpolators}

The single particle interpolating operators for $\Sigma_c$, $\bar{D}$ and $\bar{D}^*$ are:
\ba
\Sigma_{c,\alpha}^{++}&=&\epsilon^{ijk}({u^i}^T C\gamma_5 c^j) u^k_\alpha  \\
\Sigma_{c, \alpha}^{+}  &=& \frac{1}{2}\epsilon^{ijk}[({u^i}^T C\gamma_5 c^j) d^k_\alpha + ({d^i}^T C\gamma_5 c^j) u^k_\alpha] \\
D^- &=& \bar{c} \gamma_5 d, \quad \bar{D}^0 = \bar{c} \gamma_5 u \\
D_k^{*-} &=& \bar{c} \gamma_k d, \quad \bar{D}_k^{*0} = \bar{c} \gamma_k u, \quad k = 1,2,3, \
\ea
where $C$ is the charge conjugation matrix. 

The two-particle operators for $\Sigma_c \bar{D}$ and $\Sigma_c\bar{D}^*$ with $I(J^P) =  \frac{1}{2}({\frac{1}{2}}^-)$ are
\ba
\mathcal{O}^{\Sigma_c\bar{D}}_{\mathbf{p_1}, \mathbf{p_2}}  &=&\sum_{\alpha, \mathbf{p_1}, \mathbf{p_2}} C_{\alpha, \mathbf{p_1}, \mathbf{p_2}} \big( \sqrt{\frac{2}{3}} \Sigma_{c, \alpha}^{++} (\mathbf{p_{1}})D^{-} (\mathbf{p_{2}}) \nonumber \\ 
&& - \sqrt{\frac{1}{3}}\Sigma_{c, \alpha}^{+}(\mathbf{p_{1}}) \bar{D}^0(\mathbf{p_{2}}) \big),\\
\mathcal{O}^{\Sigma_c\bar{D}^*}_{\mathbf{p_1}, \mathbf{p_2}}&=& \sum_{\alpha, k, \mathbf{p_1}, \mathbf{p_2}} C_{\alpha, k, \mathbf{p_1}, \mathbf{p_2}}  \big( \sqrt{\frac{2}{3}} \Sigma_{c, \alpha}^{++}(\mathbf{p_1}) D_k^{*-}(\mathbf{p_2}) \nonumber \\ 
&& - \sqrt{\frac{1}{3}}\Sigma_{c, \alpha}^{+}(\mathbf{p_1}) \bar{D}_k^{*0}(\mathbf{p_2})\big).
\ea

We use three $\Sigma_c\bar{D}$ operators with $|\mathbf{p_{1,2}}| = 0, 1$ and $\sqrt{2}$(in units of $2\pi/L$) and two $\Sigma_c\bar{D}^*$ operators with $|\mathbf{p_{1,2}}| = 0$ and $1$. The coefficients $C_{\alpha, \mathbf{p_1}, \mathbf{p_2}}$ and $C_{\alpha, k, \mathbf{p_1}, \mathbf{p_2}}$ are chosen so that these operators transform in the $G_1^-$ irrep of the cubic group. $G_1$ is a two-dimensional representation. We use only the first row which is sufficient for the calculation. The coefficients are listed in TABLE~\ref{Table:SigmacDOps} and TABLE~\ref{Table:SigmacDstarOps} for $\Sigma_c\bar{D}$ and $\Sigma_c\bar{D}^*$ respectively. Note that these coefficients are worked out using the Dirac-Pauli representation for Dirac $\gamma$ matrices. 

\begin{table}
\begin{tabular}{|c|c|c|c|c|}
\hline
& $\alpha$ &$\mathbf{p_1}$ &$\mathbf{p_2}$ & $C_{\alpha, \mathbf{p_1}, \mathbf{p_2}}$ \\
\hline
\hline
$|\mathbf{p_{1,2}}| = 0$ & 1 & (0,0,0) & (0,0,0) & 1 \\
\hline
\hline
\multirow{6}{*}{$|\mathbf{p_{1,2}}| = 1$} &1 & (-1,0,0) & (1,0,0) & 1\\
\cline{2-5}
                                                         &1 & (1,0,0) &(-1,0,0) & 1 \\
\cline{2-5}
                                                         &1 & (0,-1,0) &(0,1,0) & 1 \\
\cline{2-5}
                                                         &1 & (0,1,0) &(0,-1,0) & 1 \\
\cline{2-5}
                                                         &1 & (0,0,-1) &(0,0,1) & 1 \\
\cline{2-5}
                                                         &1 & (0,0,1) &(0,0,-1) & 1 \\
\hline
\hline
\multirow{12}{*}{$|\mathbf{p_{1,2}}| = \sqrt{2}$} &1 & (-1,-1,0) & (1,1,0) & 1\\
\cline{2-5}
                                                                   &1 & (1,1,0) &(-1,-1,0) & 1 \\
\cline{2-5}
                                                                   &1 & (-1,0,-1) &(1,0,1) & 1 \\
\cline{2-5}
                                                                   &1 & (1,0,1) &(-1,0,-1) & 1 \\
\cline{2-5}
                                                                   &1 & (0,-1,-1) &(0,1,1) & 1 \\
\cline{2-5}
                                                                   &1 & (0,1,1) &(0,-1,-1) & 1 \\
\cline{2-5}
                                                                   &1 & (-1,1,0) &(1,-1,0) & 1\\
\cline{2-5}
                                                                   &1 & (1,-1,0) &(-1,1,0) &1 \\
\cline{2-5}
                                                                   &1 &(-1,0,1) &(1,0,-1) &1 \\
\cline{2-5}
                                                                   &1 &(1,0,-1) &(-1,0,1) &1 \\
\cline{2-5}
                                                                   &1 &(0,1,-1) &(0,-1,1) &1 \\
\cline{2-5}
                                                                   &1 &(0,-1,1) &(0,1,-1) & 1 \\
\hline
\end{tabular}
\caption{The coefficients of the $\Sigma_c\bar{D}$ operators.}
\label{Table:SigmacDOps}
\end{table}

\begin{table}
\begin{tabular}{|c|c|c|c|c|c|}
\hline
&$\alpha$ & $k$ &$\mathbf{p_1}$ &$\mathbf{p_2}$ & $C_{\alpha, k, \mathbf{p_1}, \mathbf{p_2}}$ \\
\hline
\hline
\multirow{3}{*}{$|\mathbf{p_{1,2}}| = 0$} &1  &3 &(0,0,0) & (0,0,0) & 1\\
                                                                \cline{2-6}                                                              
                                                                &2  &1 &(0,0,0) & (0,0,0) & $1$\\
                                                                \cline{2-6}
                                                                &2  &2 &(0,0,0) & (0,0,0) & $-i$\\
                                                                \hline
                                                                \hline
\multirow{6}{*}{$|\mathbf{p_{1,2}}| = 1$}  &1  &3 &(0,0,-1 & (0,0,1) & 1\\
                                                                 \cline{2-6}    
                                                                 &1  &3 &(0,0,1) & (0,0,-1) & $1$\\
                                                                 \cline{2-6} 
                                                                 &2  &1 &(1,0,0) & (-1,0,0) & $1$\\
                                                                 \cline{2-6}
                                                                 &2  &1 &(-1,0,0) & (1,0,0) & $1$\\
                                                                 \cline{2-6}
                                                                 &2  &2 &(0,-1,0) & (0,1,0) & $-i$\\
                                                                 \cline{2-6}
                                                                 &2  &2 &(0,1,0) & (0,-1,0) & $-i$\\
\hline                                                                 
\end{tabular}
\caption{The coefficients of the $\Sigma_c\bar{D}^*$ operators.}
\label{Table:SigmacDstarOps}
\end{table}                                                                 
                                                                 
\section{Computation and analysis of the correlation functions}                                                                                                                          
The distillation quark smearing method is used to compute the quark propagators. The quark smearing operator is composed of a small number($N_{ev}$) of the eigenvectors of the three-dimensional Laplacian that correspond to the $N_{ev}$ lowest eigenvalues. We compute the propagators with $N_{ev} = 100$ for the L32 ensemble and $N_{ev} = 200$ for the L48 ensemble. 

\begin{figure*}[tb]
\includegraphics[width =0.33 \textwidth]{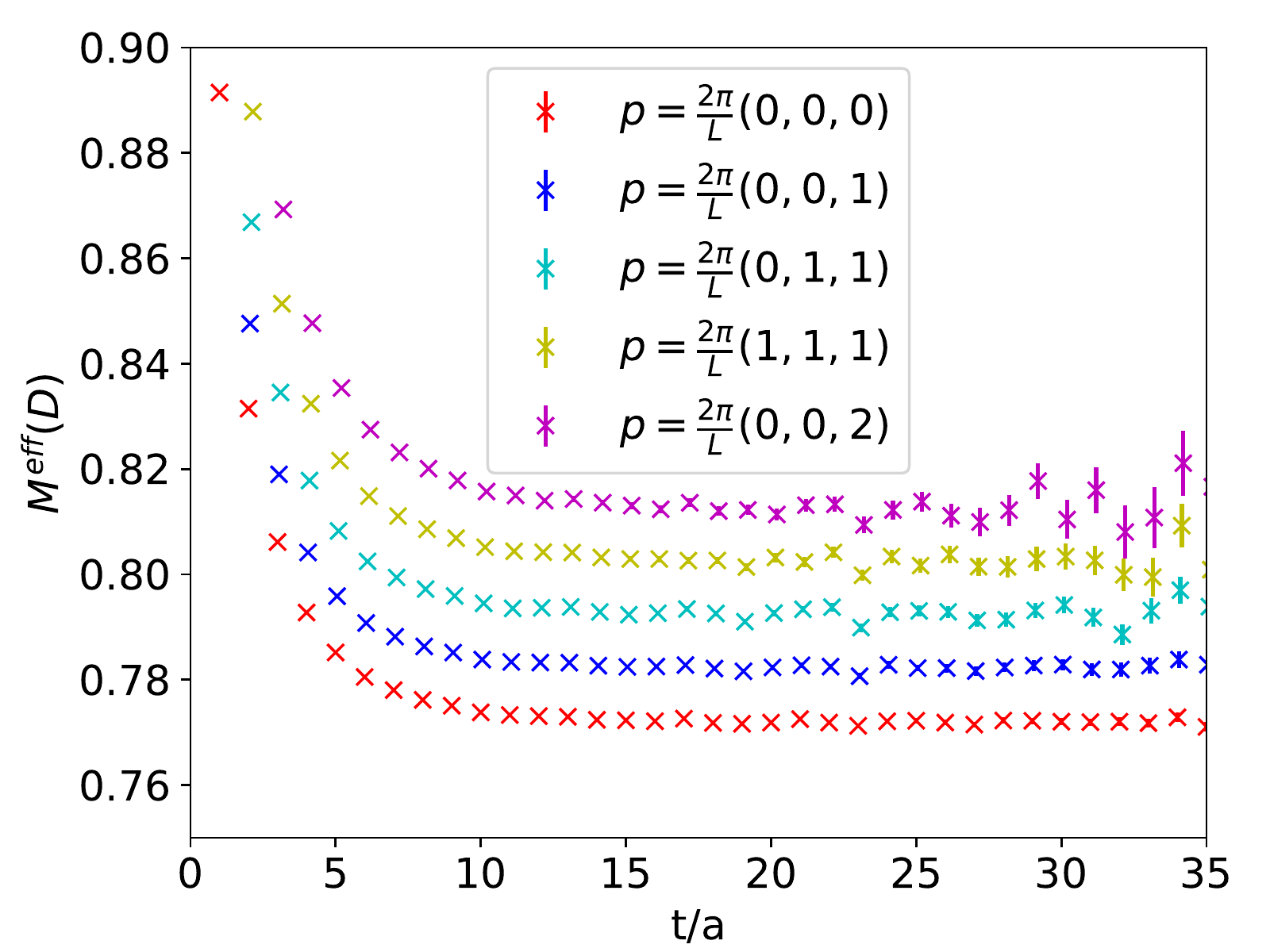}\includegraphics[width =0.33 \textwidth]{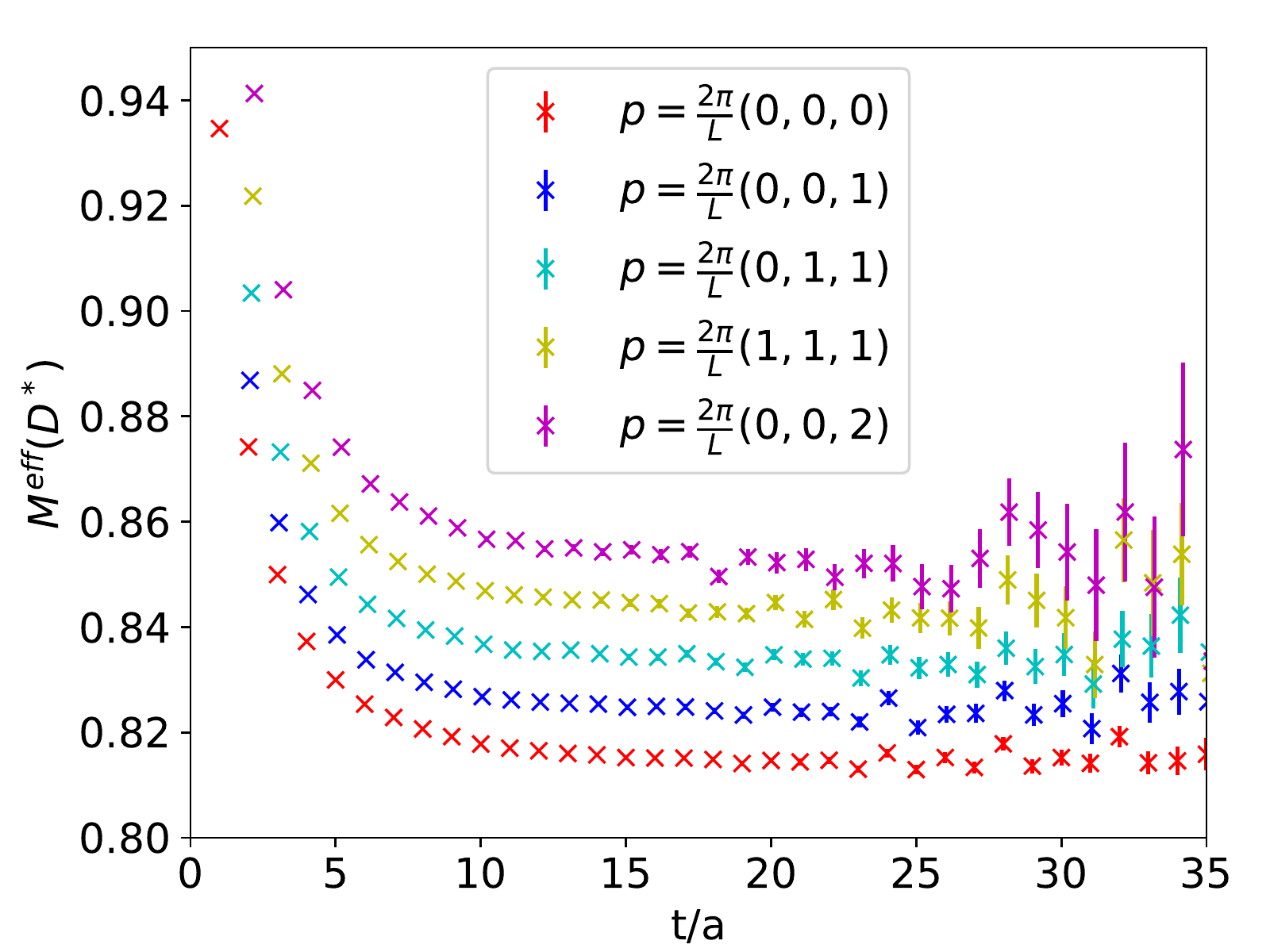}\includegraphics[width =0.33 \textwidth]{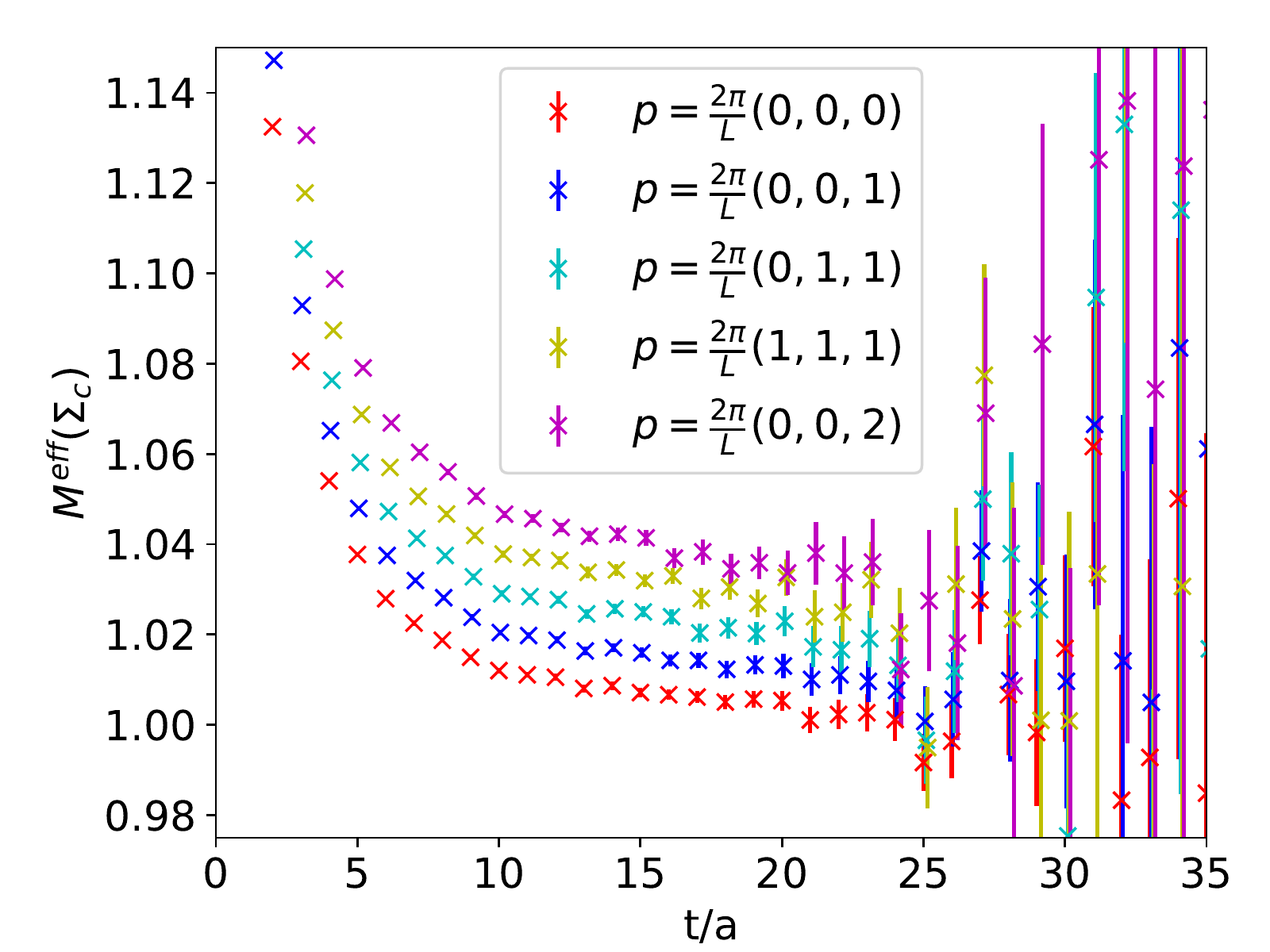}
\caption{Effective energies of $D$, $D^*$ and $\Sigma_c$ at the five lowest momenta for the ensemble L48.}
\label{Figure:em_Sigmac}
\end{figure*}

\begin{figure*}[tb]
\includegraphics[width =0.4 \textwidth]{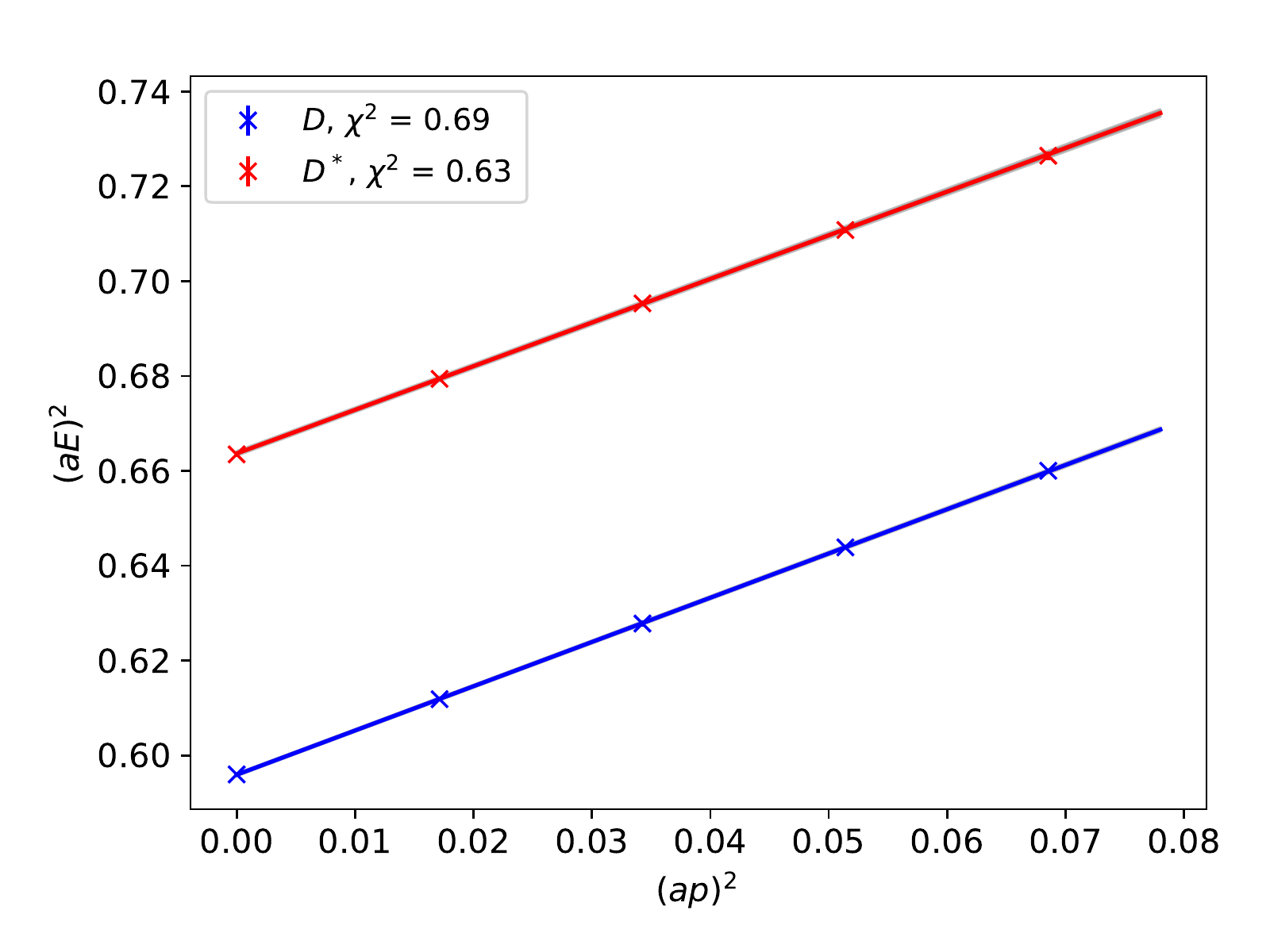}\includegraphics[width =0.4 \textwidth]{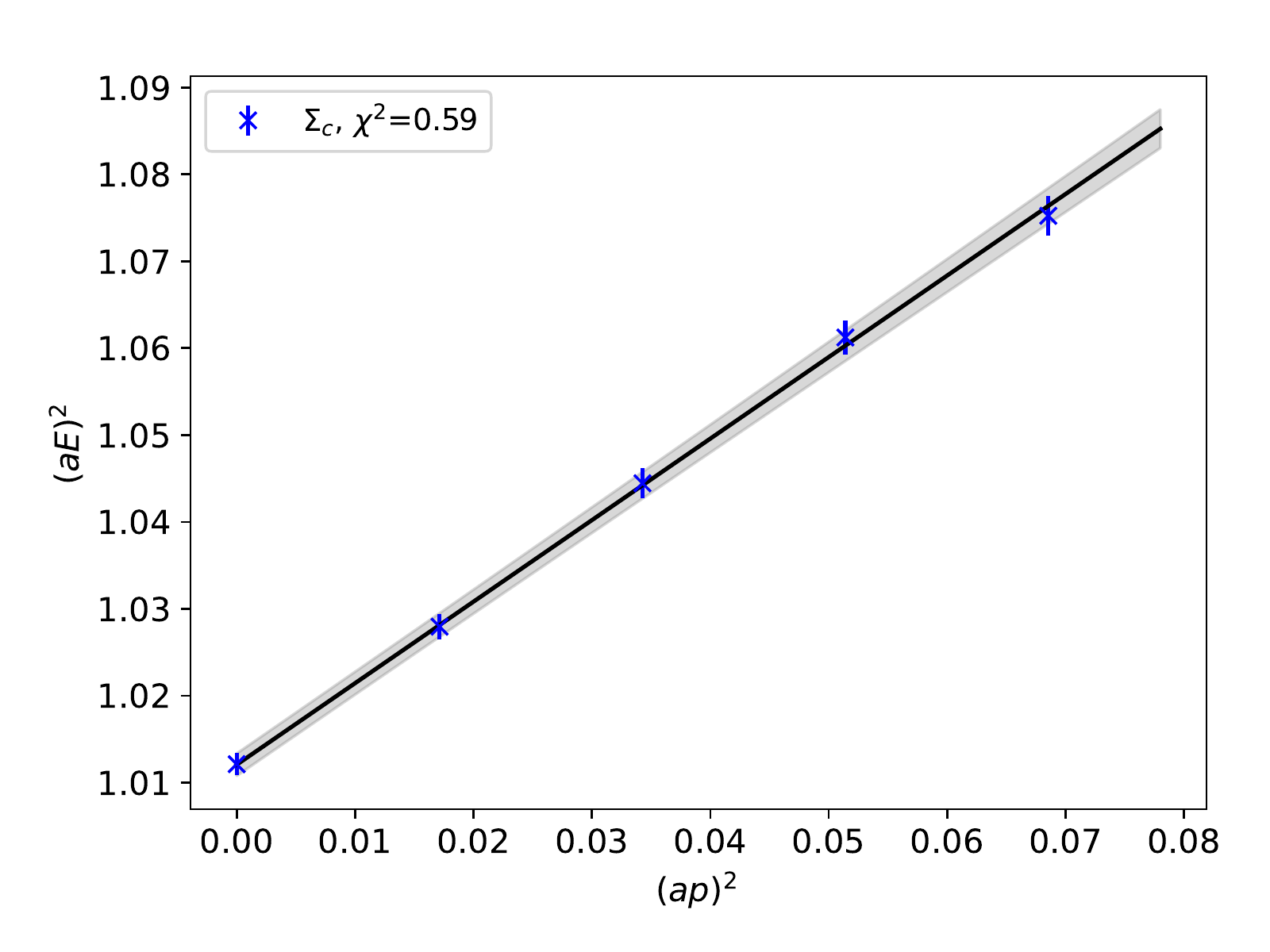}
\caption{Fits of the energies of $D$, $D^*$ and $\Sigma_c$ to the dispersion relation for the ensemble L48. The values of $\chi^2$ of the fits are shown in the plots.}
\label{Figure:Dispersion}
\end{figure*}

The single particle energies are extracted from the two-point correlation functions of the pertinent single particle operators. In FIG.~\ref{Figure:em_Sigmac}, 
we present the effective energies of $D$, $D^*$ and $\Sigma_c$ at the lowest five momenta for the ensemble L48. The fit of the five energies to the dispersion relation for each particle is shown in FIG.~\ref{Figure:Dispersion}.
 
\begin{figure*}[tb]
\includegraphics[width =0.33 \textwidth]{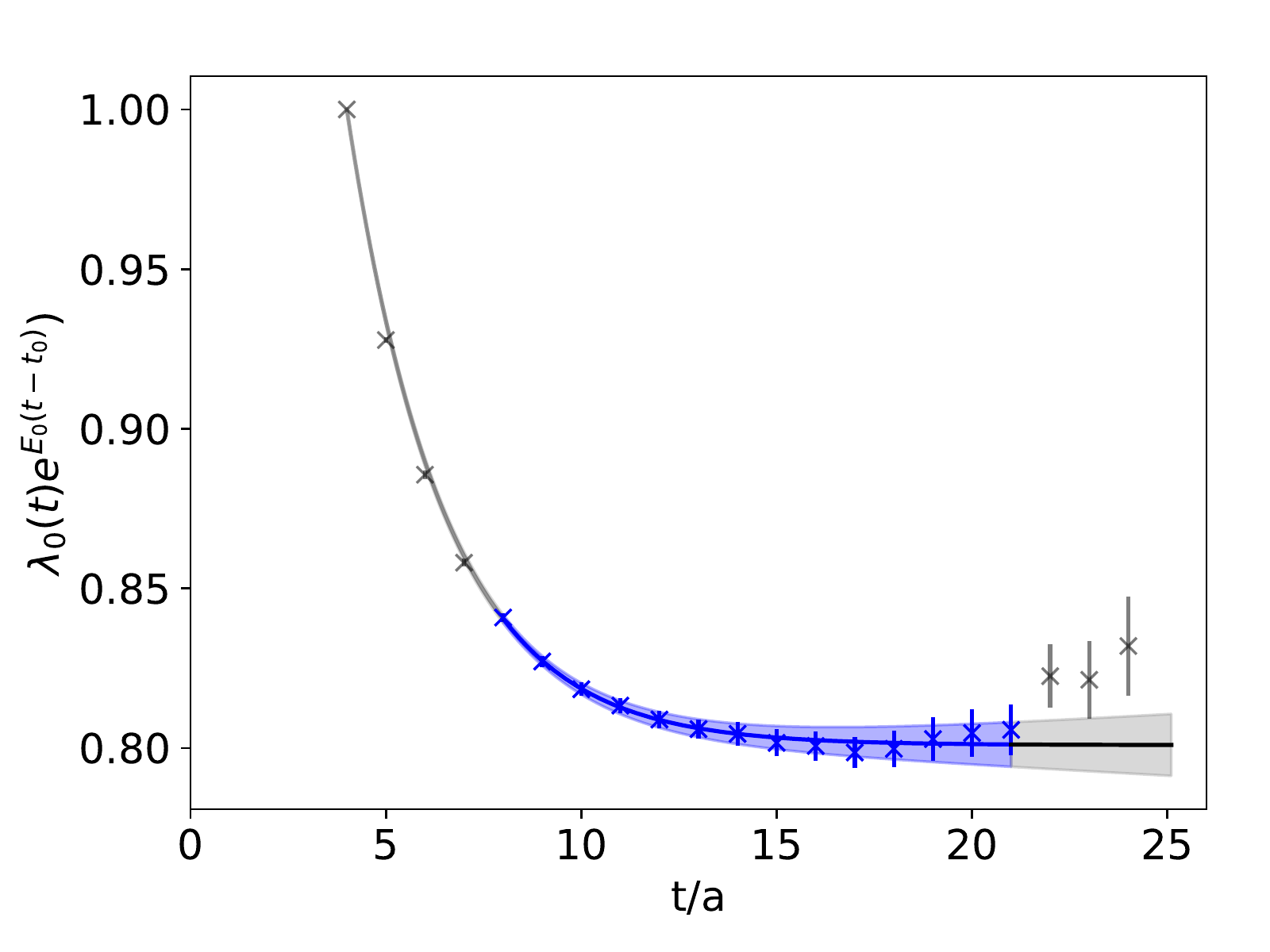}\includegraphics[width =0.33 \textwidth]{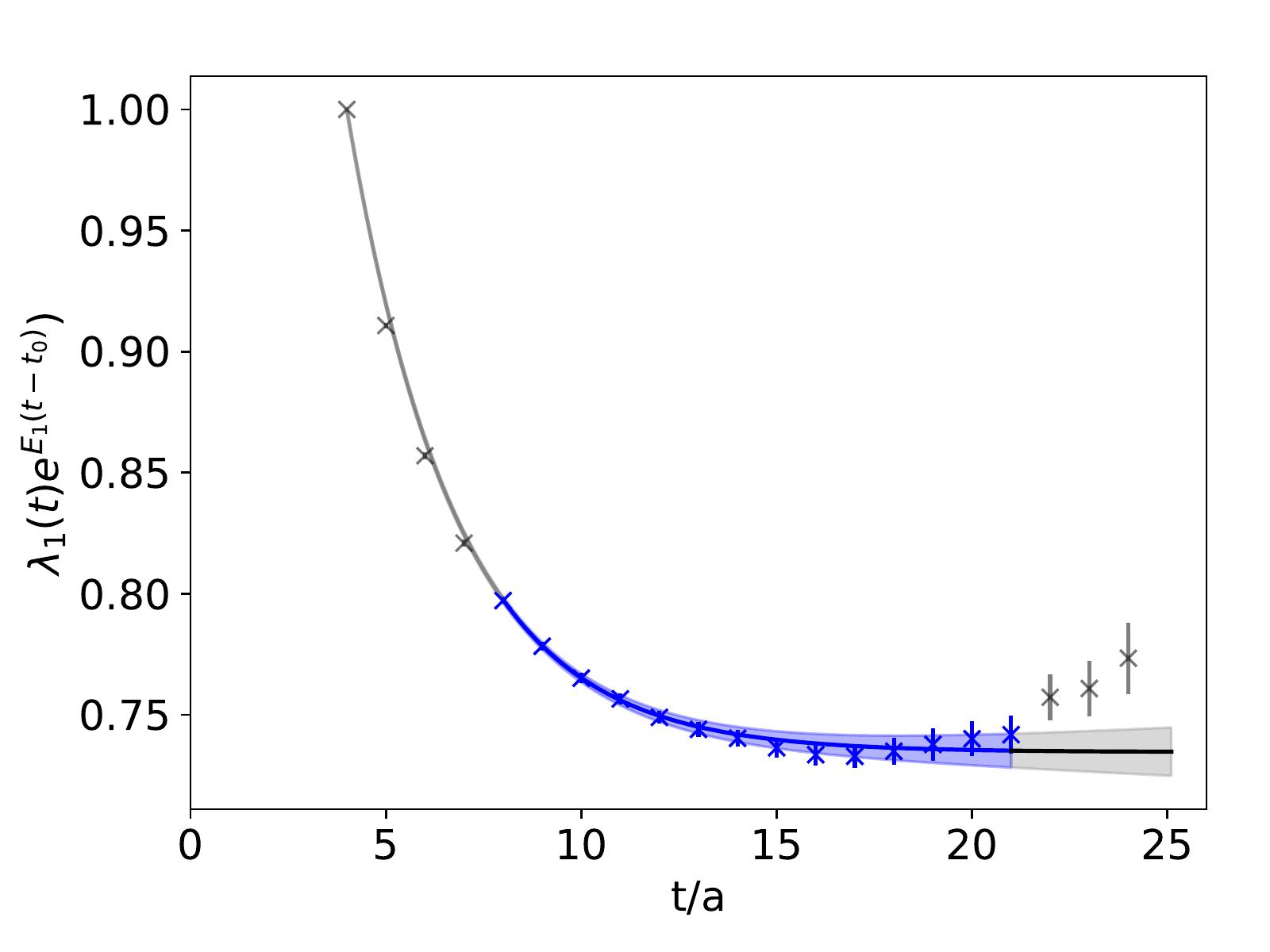}\includegraphics[width =0.33 \textwidth]{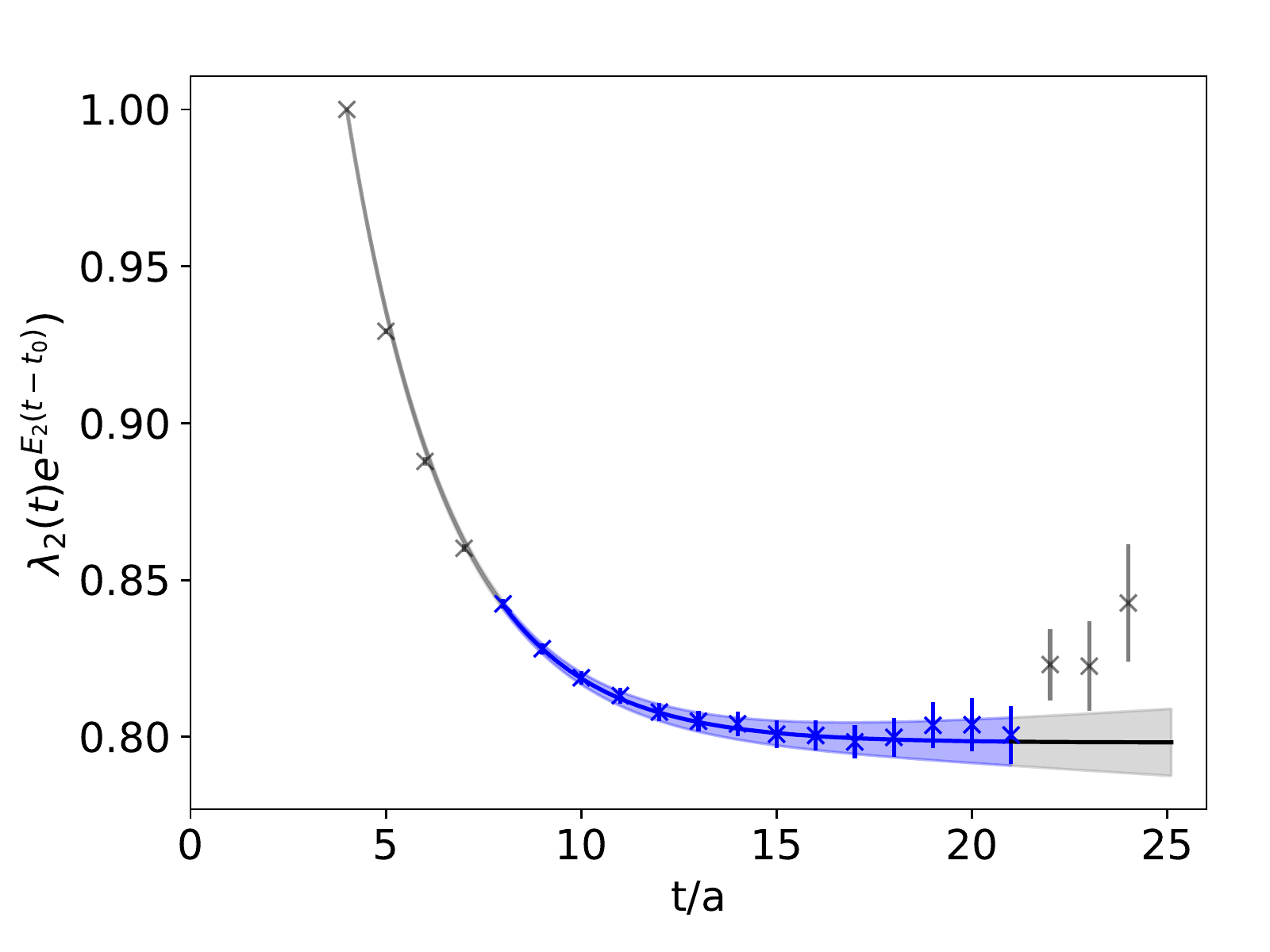}
\includegraphics[width =0.33 \textwidth]{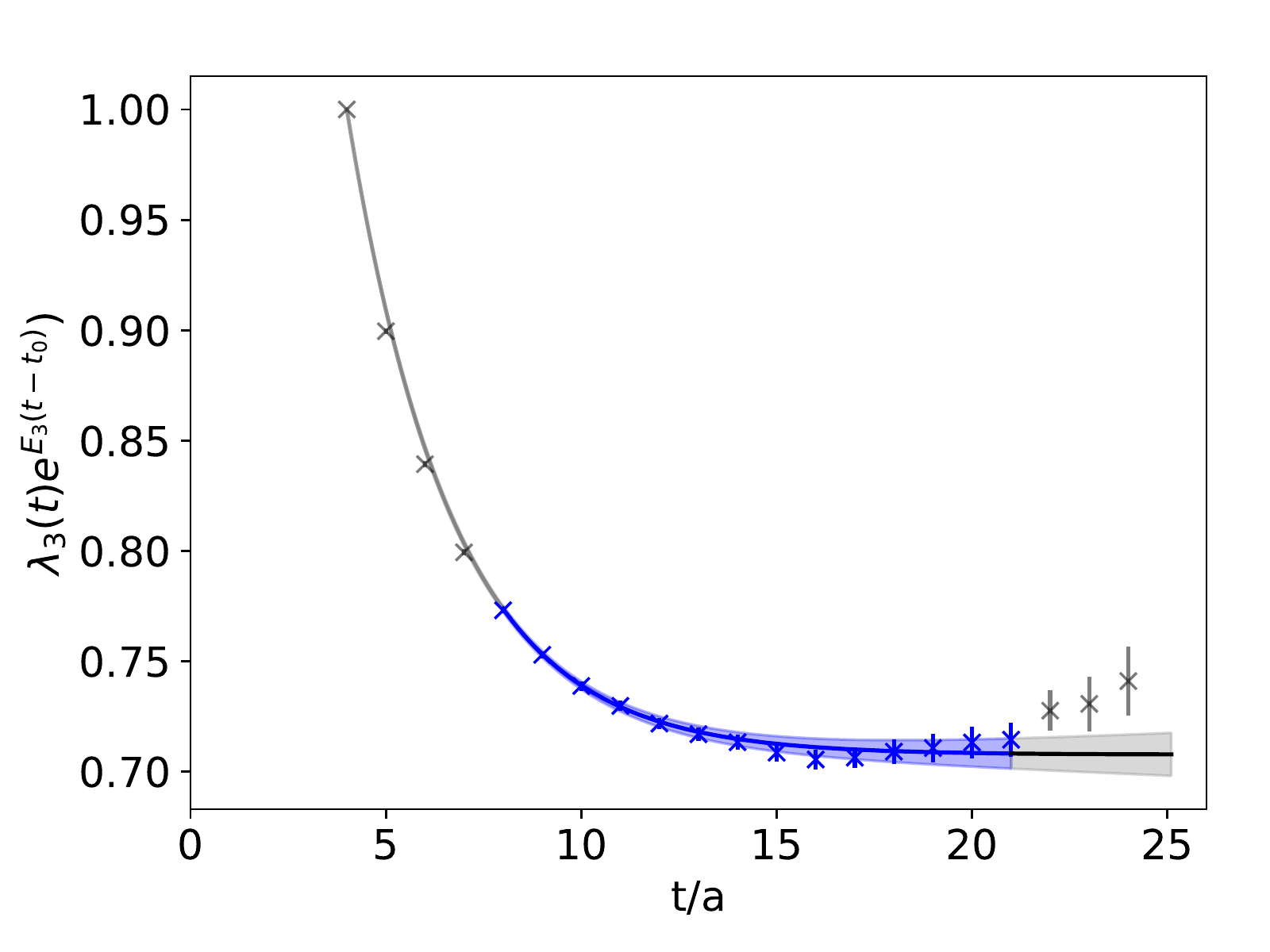}\includegraphics[width =0.33 \textwidth]{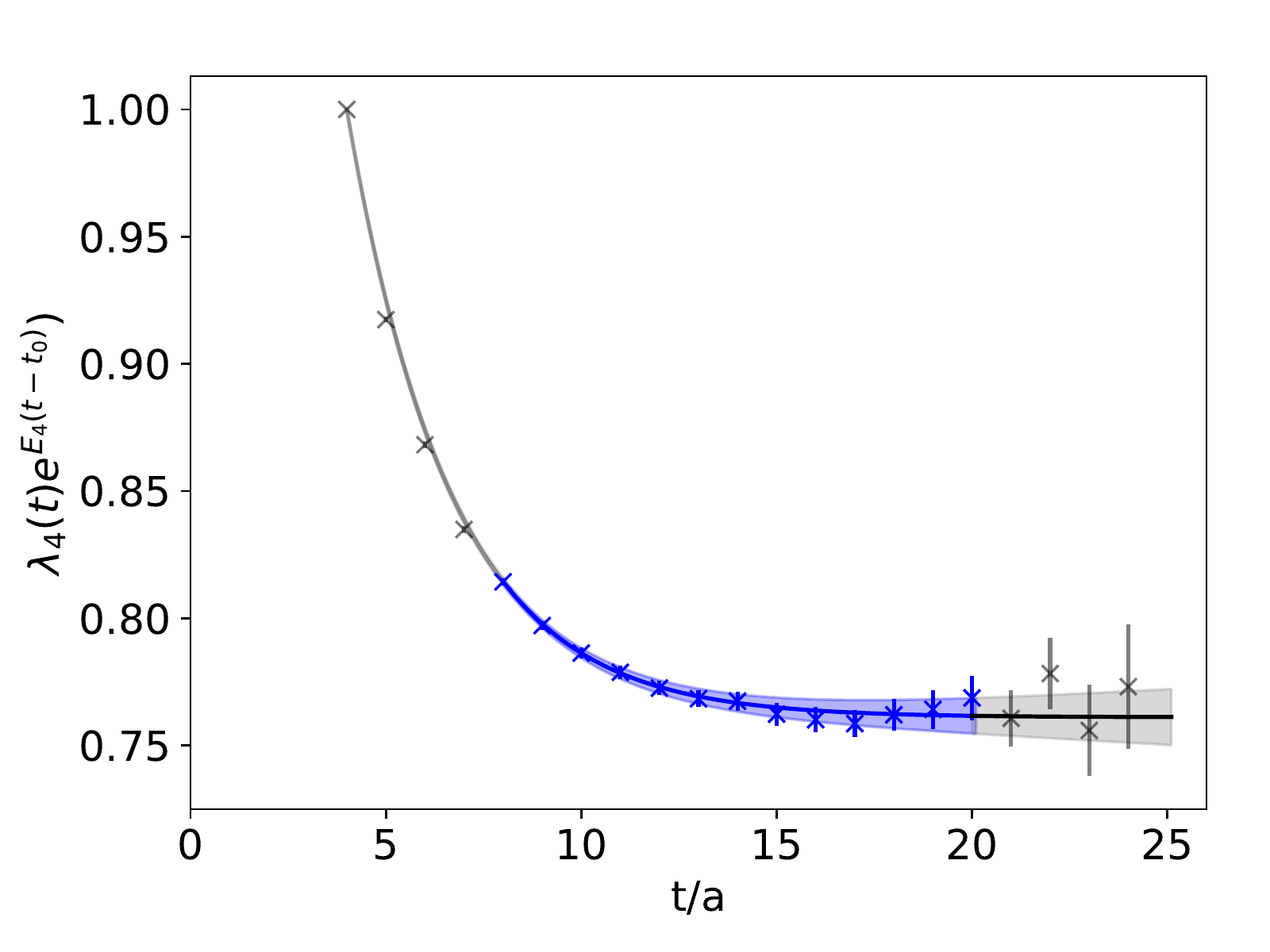}
\caption{Fits of the eigenvalues $\lambda_n(t)$. Plotted are the data $\lambda_n(t) e^{E_n(t-t_0)}$ and the fits. The blue points are those included in the fits.}
\label{Figure:Fits_Eigvals}
\end{figure*}

The finite volume two-particle energies are obtained from the matrix of the correlation functions of the five operators described in the last section. The charm quark annihilation diagrams are ignored in the calculation of the correlation functions. Solving the generalized eigenvalue problem(GEVP) 
\be
C(t) v^n(t) = \lambda^n(t) C(t_0) v^n(t),
\ee
the energies are determined by fitting the eigenvalues $\lambda^n(t)$ to the form
\be
\lambda^n(t) = (1-A_n)e^{-E_n(t-t_0)} + A_ne^{-E_n^\prime(t-t_0)},
\ee
where the fit parameters are $A_n$, $E_n$ and $E_n^\prime$. This form allows for a second exponential to capture the residual contaminations from the excited states. We tried four different values of $t_0$: 4, 6, 8 and 10, and did not observe differences in the fitted energies. The fits of the five eigenvalues for $t_0=4$ are shown in FIG.~\ref{Figure:Fits_Eigvals}
 for the ensemble L48. The fitted energies are collected in TABLE~\ref{Table:energies}
  for both ensembles. We also presented the three energies extracted from the GEVP analysis using only the $\Sigma_c\bar{D}$ operators and the two energies using only the $\Sigma_c\bar{D}^*$ operators. They agree perfectly with the values using all five operators, indicating negligible mixing between the $\Sigma_c\bar{D}$ and $\Sigma_c\bar{D}^*$ operators. 

\begin{table}
\begin{tabular}{|c|c|c|c|c|}
\hline
\multicolumn{2}{|c|}{} & all ops. & $\mathcal{O}^{\Sigma_c \bar{D}}$ & $\mathcal{O}^{\Sigma_c \bar{D}^*}$ \\
\hline
\hline
\multirow{5}{*}{L48} &$aE_0$ &1.7738(09) &1.7738(09) &1.8160(10) \\
\cline{2-5}
&$aE_1$ &1.7845(11) &1.7845(11) &1.8326(12) \\
\cline{2-5}
&$aE_2$ &1.8051(11) &1.8051(11) & -- \\
\cline{2-5}
&$aE_3$ &1.8160(10) & -- & -- \\
\cline{2-5}
&$aE_4$ &1.8326(12) &-- & -- \\
\hline
\hline
\multirow{5}{*}{L32} &$aE_0$ &1.7747(12) &1.7747(12) &1.8167(20) \\
\cline{2-5}
&$aE_1$ &1.8025(19) &1.8025(20) &1.8535(16) \\
\cline{2-5}
&$aE_2$ &1.8166(20) &1.8389(21) & -- \\
\cline{2-5}
&$aE_3$ &1.8389(21) &--  & -- \\
\cline{2-5}
&$aE_4$ &1.8535(16) &-- &-- \\
\hline
\end{tabular}
\caption{The finite volume two-particle energies. For each ensemble, we list the five energies extracted from the GEVP analysis using all five operators(all ops.). The three energies using only the $\Sigma_c \bar{D}$ operators and the two energies using only the $\Sigma_c\bar{D}^*$ operators are also presented for comparison.}
\label{Table:energies}
\end{table}

\end{document}